\documentclass[12pt]{article}
\usepackage[margin=3cm]{geometry}
\usepackage{graphicx, epstopdf, epsfig, natbib, amsmath, amssymb}
\usepackage[margin=5pt,font=small,labelfont=bf]{caption}
\usepackage{amsmath,setspace,lineno,xspace,color,tabularx,ragged2e,booktabs}
\definecolor{darkblue}{rgb}{0.0,0.0,0.6}
\usepackage[colorlinks=true,breaklinks=true,linkcolor=darkblue,
            urlcolor=darkblue,anchorcolor=darkblue,citecolor=darkblue]{hyperref} 

\newcolumntype{Y}{>{\centering\arraybackslash}X} 
\newcommand{\pdiff}[2]{\frac{\partial{#1}}{\partial{#2}}}
\newcommand{\lagdiff}[2]{\frac{\text{D}{#1}}{\text{D}{#2}}}
\newcommand{\Div}{\Grad\!\cdot}
\newcommand{\Grad}{\mbox{\boldmath $\nabla$}}
\newcommand{\delsq}{\nabla^2}
\newcommand{\strr}{\dot{e}}
\newcommand{\verysmall}{\epsilon}
\newcommand{\tausat}{\tau_\textrm{sat}}
\newcommand{\tauoff}{\tau_\textrm{offset}}
\newcommand{\anisoangle}{\Theta}
\newcommand{\compliance}{\widetilde{C}}
\newcommand{\daddtau}{\breve{\alpha}}
\newcommand{\dbddtau}{\breve{\beta}}
\newcommand{\xvec}{\boldsymbol{X}}
\newcommand{\dyfac}{\mathcal{D}}

\title{Consequences of viscous anisotropy in a deforming, two-phase
  aggregate. Why is porosity-band angle lowered by viscous
  anisotropy?}

\author{Yasuko Takei$^1$ \& Richard F.~Katz$^2$\\
  \small $^1$ Earthquake Research Institute, University of Tokyo, \\
  \small Tokyo 113-0032, Japan. $^2$ Department of Earth Sciences,
  University of Oxford, \\ \small South Parks Road, Oxford OX1 3AN,
  UK}

\begin{document}

\maketitle 
\begin{abstract}
  In laboratory experiments that impose shear deformation on partially
  molten aggregates of initially uniform porosity, melt segregates
  into high-porosity sheets (bands in cross-section).  The bands
  emerge at $15$--$20^\circ$ to the shear plane. A model of viscous
  anisotropy can explain these low angles whereas previous, simpler
  models have failed to do so.  The anisotropic model is complex,
  however, and the reason that it produces low-angle bands has not
  been understood.  Here we show that there are two mechanisms:
  (\textit{i}) suppression of the well-known tensile instability, and
  (\textit{ii}) creation of a new, shear-driven instability.  We
  elucidate these mechanisms using linearised stability analysis in a
  coordinate system that is aligned with the perturbations. We
  consider the general case of anisotropy that varies dynamically with
  deviatoric stress, but approach it by first considering uniform
  anisotropy that is imposed \textit{a priori} and showing the
  difference between static and dynamic cases. We extend the model of
  viscous anisotropy to include a strengthening in the direction of
  maximum compressive stress. Our results support the hypothesis that
  viscous anisotropy is the cause of low band-angles in experiments.
\end{abstract}

\section{Introduction}

In laboratory experiments, forced shear deformation of nominally
uniform, partially molten rocks causes melt segregation into
high-porosity bands oriented at low angle ($15$--$20^\circ$) to the
shear plane \citep{holtzman03a, holtzman07, king10}.
\cite{stevenson89} predicted the emergence of such bands in a
self-reinforcing feedback arising from the porosity-weakening of the
crystal$+$magma aggregate, but the angle predicted by this theory was
$45^\circ$ \citep{spiegelman03b}, much higher than observed.  The low
angle of high-porosity bands is widely thought to provide an
additional constraint on the rheology of the aggregate, but it has
proven challenging to understand.  \cite{katz06} found that
non-Newtonian viscosity with a high sensitivity to stress could
reproduce the low angle of bands, but \cite{king10} subsequently
showed that the viscosity of experiments that produce low-angle bands
is actually close to Newtonian.

Theory by \cite{takei09a, takei09c} of anisotropic viscosity under
diffusion creep of a partially molten aggregate represents a possible
solution. This theory is motivated by observations of the coherent
alignment of melt-pockets between solid grains under a deviatoric
stress \citep[e.g.][]{daines97, zimmerman99} and of the enhancement of
diffusion creep by melt at grain boundaries and triple junctions
\citep[e.g.][]{cooper89}.  The melt is a fast pathways for diffusional
transport of solid constituents around grains; the alignment of melt
with respect to the principal-stress directions 
hypothetically results in anisotropic viscosity of the aggregate
\citep{takei09a}.

Analysis of the theory of anisotropic viscosity by \cite{takei09c},
 \cite{takei13}, \cite{katz13}, and \cite{allwright14}
shows that it introduces qualitatively different behaviour from
previous models with isotropic (and even power-law) viscosity.  Shear
and normal components of stress and strain-rate are coupled under
viscous anisotropy; as a result of this coupling, a gradient in shear
stress becomes a driving force for melt segregation that is not
present in the isotropic system. Under Poiseuille flow, melt
segregates toward higher-stress regions; under torsional flow,
compressive hoop stresses drive the solid outward and the magma
inward. The mechanics of this ``base-state'' melt segregation are
explained in detail by \cite{takei13}. An experimental test of radial
melt segregation in torsional flow by \cite{qi15} shows striking
consistency with predictions.

Furthermore, theoretical work has demonstrated that there is a
connection between the strength of anisotropy and the angle of
high-porosity bands that emerge by unstable growth.  This was shown
with linearised stability analysis \citep{takei09c, takei13} and numerical
simulations \citep{butler12, katz13} where the strength and orientation of
anisotropy are assumed to be known and are imposed \textit{a priori}.
In those static-anisotropy calculations, high-porosity bands emerge at
low angles to the shear plane only when viscous anisotropy is at or
near saturation.  This is a rather restrictive condition that may be
incompatible with the robust appearance and consistently low angle of
bands in experiments \citep{holtzman07}. However, in numerical
simulations that allow anisotropy strength and direction to vary
dynamically in space and time \citep{katz13}, band angles are
significantly lowered and appear to be less sensitive to the mean
strength of anisotropy.  These findings raise several basic,
unanswered questions: Why do the mechanics of viscous anisotropy give
rise to low-angle bands?  Why is dynamic anisotropy more effective in
this regard than static anisotropy?  What are the general conditions
under which low-angle, high-porosity bands should form?

The present manuscript addresses these questions through a combination
of linearised stability analysis and physical reasoning.  The crucial,
enabling advance is to perform the analysis in a coordinate system
that is rotated to align with the porosity bands (rather than with the
plane of shear).  This drastically simplifies the expressions for
growth rate under static anisotropy \citep{takei13}, making them
readily interpretable in physical terms.  Moreover, it allows us to
extend the analysis to dynamic anisotropy in a form that exposes the
physical differences from static anisotropy.  Finally, the same
coordinate rotation clarifies the physical reason for low angles under
isotropic, non-Newtonian viscosity.

The manuscript is organised as follows.  In the next section, we
briefly discuss the nondimensionalised governing equations and present
an anisotropic, viscous constitutive model for the two-phase,
partially molten aggregate.  The full, non-linear system is solved
numerically in \S{\ref{sec:numerical}} for static and dynamic cases,
to elucidate the questions listed above.  The coordinate rotation is
introduced and the linearised stability analysis is developed in
\S{\ref{sec:linear}}. In particular, \S{\ref{sec:general-solution}}
develops an expression for the growth-rate of porosity perturbations
under the fully dynamic model of \S{\ref{sec:governing-eqns}}.  This
expression is challenging to understand and so we subsequently
consider it under reducing assumptions of static anisotropy
(\S{\ref{sec:static}}), which includes the simplest case of Newtonian,
isotropic model.  We build on this to explain the full complexity in
\S{\ref{sec:dynamic}} and \S{\ref{sec:beta}}.  We conclude with a
summary and discussion of the results in terms of the motivating
questions.

\section{Governing and constitutive equations}
\label{sec:governing-eqns}

In the theory of magma/mantle interaction, the macroscopic behaviour
of a two-phase aggregate is treated within the framework of continuum
mechanics \citep[e.g.][]{drew83, mckenzie84}. This theory is concerned
with the evolution of macroscopic fields including the volume fraction
of melt or porosity $\phi$, the velocity of the solid phase
$\mathbf{V}$, the liquid pressure $P$ (compression positive), and the
bulk or phase-averaged stress tensor
$\sigma_{ij} = (1-\phi)\sigma^S_{ij} - \phi P \delta_{ij}$ where
$\sigma^S_{ij}$ is the stress tensor of the solid phase (tension
positive). Further details of the two-phase-flow theory were
previously presented \citep[e.g.][]{takei13, rudge11} and are not
repeated here.

We proceed directly to the nondimensional governing equations,
\begin{subequations}
  \label{eq:governing}
  \begin{align}
    \pdiff{\phi}{t}
    &=\Div \left[ (1-\phi) {\bf V}\right], \label{eq:eqmo1}\\
    \Div {\bf V}
    &=\frac{R^2}{r_\xi+4/3}  
      \Div \left[ \left(\frac{\phi}{\phi_0}\right)^\ell 
      \Grad P \right], \label{eq:eqmo2}\\
    \Grad P  &= \Div\boldsymbol{\tau}, \label{eq:eqmo3}
  \end{align}
\end{subequations}
and refer the reader to \cite{takei13} and references therein for
details of the derivation and rescaling. In the system
\eqref{eq:governing} we have introduced the differential stress
tensor $\tau_{ij}\equiv\sigma_{ij}+P\delta_{ij}$. Also, we have
excluded body forces and assumed that the permeability of the solid
matrix is a function of the porosity only, proportional to
$(\phi/\phi_0)^\ell$, where $\phi_0$ is a reference porosity and
$\ell$ is a constant. $R$ is the nondimensional compaction length and
$r_\xi$ is a rheological parameter explained below. To close the
system, a constitutive relationship that relates the differential
stress $ \tau_{ij} $ and strain rate $\strr_{ij}=(V_{i,j} +
V_{j,i})/2$ is required.  

\cite{takei09c} and \cite{takei13} proposed a model of anisotropic
viscosity caused by stress-induced microstructural anisotropy.  In
partially molten rocks, the melt phase is contained within a permeable
network of tubules between grains. The solid matrix is formed by a
contiguous skeleton of solid grains. The area of grain-to-grain
contact is known as the contiguity. Contiguity is the microstructural
variable that determines the macroscopic (i.e., continuum) mechanical
properties of the matrix \citep{takei98, takei09a}.  Although the
equilibrium microstructure developed under hydrostatic stress has
isotropic contiguity, deviations from the equilibrium microstructure
have been observed in experimentally deformed, partially molten
samples \citep[e.g.][]{daines97, takei10}. Based on these
observations, we infer that under a differential stress, the
grain-to-grain contacts with normals that are parallel to the maximum
tensile stress ($\tau_3$) are reduced in area; similarly, the areas of
those with normals parallel to the maximum compressive stress
($\tau_1$) are increased.  Using a microstructure-based model of
aggregate viscosity \citep{takei09a} and a coordinate transformation
\citep{takei13}, the constitutive law and the viscosity tensor are
\begin{subequations}
  \label{eq:constitutive}
  \begin{equation}
    \tau_{ij} = C_{ijkl}\strr_{kl},
  \end{equation}
  \begin{multline}
    C_{ijkl}  = \textrm{e}^{-\lambda(\phi-\phi_0)} \times \\
    \bordermatrix{ ij\!\downarrow &kl\!\rightarrow XX&YY&XY  \\[1mm]
      XX&r_\xi+\frac{4 }{3} -\frac{\alpha+\beta}{2} \cos2\anisoangle 
      & r_\xi-\frac{2}{3} &-\frac{\alpha+\beta}{4}  \sin2\anisoangle \\[-1mm]
      &-\frac{\alpha-\beta}{8} (3+ \cos4\anisoangle)
      &-\frac{\alpha-\beta}{8} (1- \cos4\anisoangle) &
      -\frac{\alpha-\beta}{8} \sin4\anisoangle   \\[2mm]
      YY& \cdot  & r_\xi+\frac{4 }{3}+\frac{\alpha+\beta}{2}  
      \cos2\anisoangle &-\frac{\alpha+\beta}{4}  \sin2\anisoangle   \\[-1mm]
      &    &-\frac{\alpha-\beta}{8} (3+ \cos4\anisoangle)
      &+\frac{\alpha-\beta}{8} \sin4\anisoangle   \\[2mm]
      XY&\cdot &\cdot&1-\frac{\alpha-\beta}{8} (1- \cos4\anisoangle)}.
    \label{eq:Csystem}
  \end{multline}
\end{subequations}
For simplicity, we consider a two dimensional problem, in which the
$\tau_1$--$\tau_3$ plane is parallel to the $X$--$Y$ plane. Therefore
only the two-dimensional version of $C_{ijkl}$ is written in
\eqref{eq:Csystem}. Only 6 of the 16 components are shown due to the
symmetry of $C_{ijkl}$ under the exchange of $i$ and $j$, $k$ and $l$,
and $ij$ and $kl$.

The factor in front of the matrix represents the normalised shear
viscosity $\eta(\phi)/\eta(\phi_0)$; it decreases exponentially with
increasing melt fraction $\phi$, and so $\lambda$ is called the
porosity-weakening factor. We take $\lambda = 27$ based on the
experimental results \citep[e.g.][]{mei02}.  The parameter $r_\xi$
represents the bulk-to-shear viscosity ratio, $r_\xi=\xi / \eta$,
which is assumed to be constant (=5/3) based on theoretical results by
\cite{takei09a} \citep[although see][]{simpson10a, simpson10b}.
Parameters $\alpha$, $\beta$, and $\anisoangle$ represent the
magnitude and direction of microstructural anisotropy: $\alpha$ and
$\beta$ quantify the amplitude of contiguity reduction and increase,
respectively, in the directions of principal stress $\tau_3$ and
$\tau_1$; $\anisoangle$ represents the angle that the most-tensile
stress ($\tau_3$) direction makes with the $X$-axis of the coordinate
system. Using the local differential stress $\tau_{ij}$, $\anisoangle$
is given by
\begin{equation}
  \tan 2\anisoangle  = \frac{2\tau_{XY}}{\tau_{XX}-\tau_{YY}},
  \label{eq:Theta}
\end{equation}
and $\alpha$ and $\beta$ are modelled as 
\begin{subequations}
  \label{eq:alphabeta}
  \begin{align}
    \label{eq:alpha_dynamic}
    \alpha&= 1+ \tanh \displaystyle\left(
            \frac{2(\Delta \tau-\tauoff)}{\tausat}
            \right),\\
    \label{eq:beta_dynamic}
    \beta&= r_\beta \alpha,
  \end{align}
\end{subequations}
where $\Delta \tau$ $=\tau_3-\tau_1$
$ =\sqrt{(\tau_{XX}-\tau_{YY})^2+4\tau_{XY}^{2}}$ represents the
amplitude of deviatoric stress.  The detailed forms of the functions
in \eqref{eq:alphabeta} are poorly constrained, owing to a lack of
experimental data. The form of $\alpha$ was chosen based on the
constraints that $\alpha$ is less than or equal to 2 \citep{takei13}
and increases with increasing differential stress \citep{daines97,
  takei10}.  The parameter $r_\beta$ is assumed to be a constant that
is probably between 0 and 1.  In the present study, $\alpha$ is
parameterised by $\tauoff$ and $\tausat$, which control the
stress-offset and slope of increase.  In \cite{takei13}, $\beta=0$ and
$\alpha$ was parameterised by $\tausat$ alone. Parameters $\beta$ and
$\tauoff$ are newly introduced here.

For simplicity in previous linearised analyses \citep{takei13,
  allwright14}, parameters $\alpha$, $\beta$, and $\anisoangle$ were
fixed to their initial values. We call this simplifying assumption the
\textit{static anisotropy} model. In contrast, the complete model with
stress-dependent direction and magnitude is called the \textit{dynamic
  anisotropy} model. \cite{katz13} discovered a remarkable difference
between static and dynamic anisotropy; this difference motivates the
present study and is demonstrated in the next section.

\section{Numerical solutions}
\label{sec:numerical}

Numerical solutions of equations \eqref{eq:governing} and
\eqref{eq:constitutive} highlight the difference between results
obtained for static and dynamic anisotropy.  The solutions are
computed with a finite-volume method on a fully staggered grid that is
periodic in the $X$-direction; in this section, the $X$ axis is taken
parallel to the initial flow direction. No-slip, impermeable boundary
conditions enforce a constant displacement rate of plus or minus
$\tfrac{1}{2}\hat{\xvec}$ on the top and bottom boundaries,
respectively. A semi-implicit, Crank-Nicolson scheme is used to
discretise time and the hyperbolic equation for porosity evolution is
solved separately from the elliptic system in a Picard loop with two
iterations at each time-step. The solutions are obtained in the
context of the Portable, Extensible Toolkit for Scientific Computation
\citep[PETSc,][]{petsc-homepage, petsc-manual, katz07}.  Full details
and references are provided by \cite{katz13}.

\begin{figure}
  \centering \vspace{3mm}
  \includegraphics[width=\textwidth]{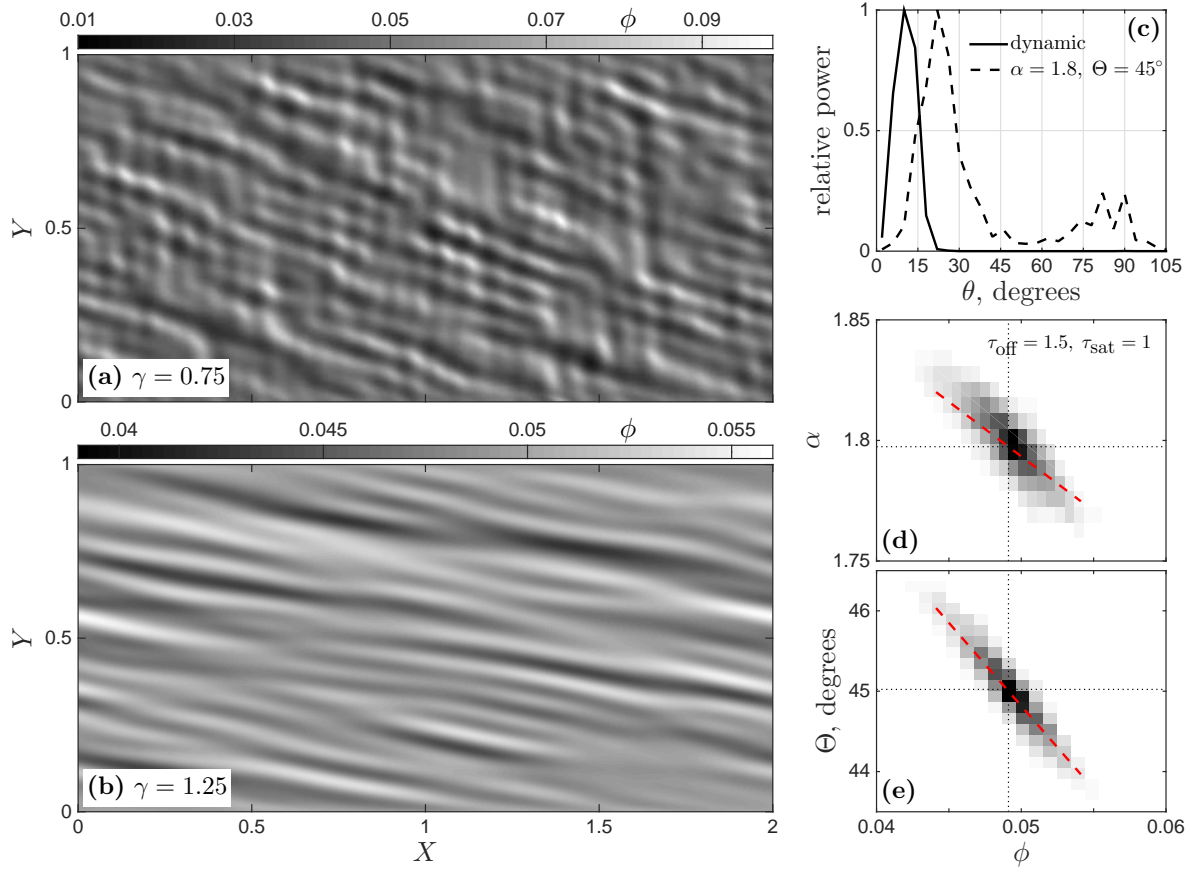}
  \caption{Comparison of numerical solutions to \eqref{eq:governing}
    and \eqref{eq:constitutive} with static and dynamic
    anisotropy. For both calculations,
    $r_\beta=1,\,R=1,\,r_\xi=5/3,\,\phi_0=0.05,\,
    \epsilon\vert\phi_1(\xvec)\vert\le0.005$
    and the domain is discretised into 600$\times$300 square
    cells. \textbf{(a)} Porosity field at a strain of 0.75 for a
    simulation with $\alpha=1.8$ and $\Theta=45^\circ$ throughout the
    domain. \textbf{(b)} Porosity field at a strain of 1.25 for a
    simulation with anisotropy calculated according to
    \eqref{eq:Theta} and \eqref{eq:alphabeta} with
    $\tausat=1,\,\tauoff=1.5$. \textbf{(c)} Spectral power binned by
    wavefront-angle $\theta$ to the shear plane
    \citep[after][]{katz06} for the porosity fields shown in panels
    (a) and (b). Each spectrum is normalised by its maximum
    power. \textbf{(d)} and \textbf{(e)} Two-dimensional histograms
    derived from the simulation with dynamic anisotropy at a strain of
    1.25 \citep[after][Fig.~12]{katz13}. Red dashed lines have a slope
    given by the ratio of perturbation quantities $\alpha_1/\phi_1$
    and $\Theta_1/\phi_1$ from the stability analysis in
    \S{\ref{sec:linear}}.}
  \label{fig:numerical}
\end{figure}

Figure \ref{fig:numerical} compares solutions with fixed and dynamic
anisotropy.  In panel~(a), anisotropy parameters are prescribed as
$\alpha=\beta=1.8$, $\Theta=45^\circ$; in panel~(b), these parameters
are calculated cell-wise using equations~\eqref{eq:Theta} and
\eqref{eq:alphabeta}, with $r_\beta=1$. Both calculations have $R=1$
(compaction length equal to domain height) and are initialised with
the same porosity field,
$\phi(\xvec,t=0) = \phi_0 + \verysmall\phi_1(\xvec)$, where
$\phi_0=0.05$ and $\epsilon=0.005$. $\phi_1(\xvec)$ is a smooth,
random field with unit amplitude, generated by filtering grid-scale
white noise to remove variation at wavelengths below 15
grid-cells. Because the growth-rate of porosity perturbations differs
for fixed and dynamic anisotropy, the simulations are shown at
different values of the average simple-shear strain $\gamma$.

The different orientation of high-porosity features is evident in
panels~(a) and (b): dynamic anisotropy is associated with lower
angles. This is quantified by the power spectrum in panel~(c), where
the power from a 2D fast-Fourier transform of the porosity field is
binned according to the angle between the wavefront and the shear
plane \citep{katz06}. Dynamic anisotropy produces a peak at
$\sim10^\circ$ whereas static anisotropy produces a peak at
$\sim23^\circ$.  There is also a high-angle ($\sim80^\circ$) peak for
static anisotropy (corresponding to features visible in panel~(a))
that does not survive at large strain.  Panels~(d) and (e) show the
covariation of $\alpha$ and $\Theta$ with $\phi$ in panel~(b); black
dotted lines indicate mean values. These means are closely matched
with the parameter values used in the fixed-anisotropy simulation.  It
is therefore clear that the difference in the dominant band angle
(panel~(c)) arises from the coupling between stress and the variations
in $\alpha$ and $\Theta$.  What is unclear, however, is the physical
explanation for this difference and, indeed, why viscous anisotropy
gives rise to bands at angles less than $45^\circ$ to the shear plane
at all.  We clarify these points below.

\section{Linearised analysis with a perturbation-oriented coordinate
  system}
\label{sec:linear}

Let $\verysmall\ll\phi_0$ be the initial amplitude of a porosity
perturbation. We express the problem variables as a series expansion
about the base-state in which the porosity is uniform and equal to
$\phi_0$. We truncate the series after the first-order terms,
\begin{equation}
  \left\{
    \begin{array}{ccccc}
      \phi (\xvec,t)&=&\phi_0 &+&\verysmall \phi_1(\xvec, t)\\
      P (\xvec,t)&=&0 &+&\verysmall P_1(\xvec, t)\\
      {\bf V} (\xvec,t) &=& {\bf V}^{(0)} (\xvec)&+&\verysmall {\bf V}^{(1)} (\xvec,t)\\
      \strr_{ij} (\xvec,t)&=& \strr^{(0)}_{ij}    &+&\verysmall
      \strr^{(1)}_{ij}(\xvec, t)\\
      \tau_{ij} (\xvec,t)&=& \tau^{(0)}_{ij}    &+&\verysmall \tau^{(1)}_{ij}(\xvec, t)\\
      C_{ijkl} (\xvec,t)&=&C^{(0)}_{ijkl}&+&\verysmall C^{(1)}_{ijkl}(\xvec, t)\\
      \alpha (\xvec,t) &=& \alpha_0 &+& \verysmall \alpha_1
      (\xvec,t) \\
      \anisoangle (\xvec,t) &=& \anisoangle_0 &+& \verysmall \anisoangle_1  (\xvec,t). \\
    \end{array}
  \right.
  \label{eq:expand}
\end{equation}
The first term of \eqref{eq:expand} with index $0$ represents a
simple-shear flow and its associated anisotropy, which is the
base-state solution of order one ($\verysmall^0$), corresponding to
the uniform porosity $\phi_0$.  The second term of \eqref{eq:expand}
with index 1 represents the perturbation of order $\verysmall^1$
caused by $\verysmall\phi_1$.  By substituting \eqref{eq:expand} into
equations~\eqref{eq:governing}, using $\Div{\bf V}^{(0)}=0$, and
balancing terms at the order of $\verysmall^1$, we derive the
governing equations for the perturbations as
\begin{subequations}
  \label{eq:first-order}
  \begin{align}
    \lagdiff{\phi_1}{t}  
    &= (1-\phi_0)\Div {\bf V}^{(1)} \label{eq:eqmo1_1st}\\
    \Div{\bf V}^{(1)} 
    &= \displaystyle\frac{R^2}{r_{\xi}+4/3}\delsq  P_1 \label{eq:eqmo2_1st}\\
    P_{1,i} &= \left[C^{(1)}_{ijkl}\strr^{(0)}_{kl}\right]_{,j} +
              \left[C^{(0)}_{ijkl}\strr^{(1)}_{kl}\right]_{,j}
              \,\, (=\tau^{(1)}_{ij,j}) ,
              \label{eq:eqmo3_1st}
  \end{align}
\end{subequations}
where $\text{D} \phi_1/ \text{D} t=\partial \phi_1 / \partial t+ {\bf V}^{(0)}
\cdot \Grad \phi_1$.  

\begin{figure}
  \centering  \vspace{3mm}
  \includegraphics[width=\textwidth]{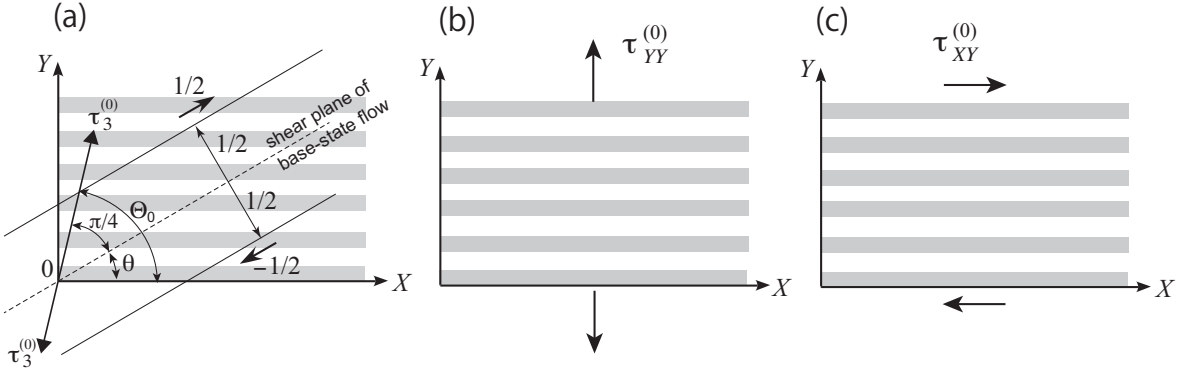}
  \caption{Schematic diagrams of the coordinate axes and porosity
    perturbation. \textbf{(a)} The coordinate system $(X,Y)$ for the
    linearised analysis is taken such that the $X$ axis is parallel to
    the initial perturbation wavefronts. The shear plane of the
    base-state, simple-shear flow is then rotated by an angle
    $\theta$. \textbf{(b)} The base-state normal stress
    $\tau^{(0)}_{YY}$ is oriented parallel to the initial perturbation
    wave vector. \textbf{(c)} The base-state shear stress
    $\tau^{(0)}_{XY}$ is parallel to the wavefronts.}
  \label{fig:schematic}
\end{figure}

Following previous studies, the porosity perturbations $\phi_1$ take
the form of a plane wave oriented at a given angle to the base-state
shear plane.  Past workers chose to align the coordinate system with
the base-state shear plane, such that the base-state strain rate
tensor has a simple form \citep[e.g.][]{spiegelman03b}.  Although the
coordinate system was so aligned in the numerical models above, in
this section the coordinates are rotated such that the $Y$ axis is
parallel to the wave-vector of the initial perturbation, as shown in
Figure~\ref{fig:schematic}. With this choice, $\theta$ again
represents the angle between the perturbation wavefronts and the
base-state shear plane.  However, in the
rotated coordinate system, $\Theta$ depends on both the direction of
$\tau_3$ and the orientation of the bands.  The base-state direction
of maximum tensile stress $\tau_3$ makes an angle $\pi/4$ to the shear
plane (eqn.~\eqref{eq:Theta}) and so for a coordinate rotation by
$\theta$, we have $\Theta_0 = \pi/4 + \theta$
(Figure~\ref{fig:schematic}a).

In the following part of this section, we give an outline of the
linearised approach, which shows that the new coordinate system
reduces the complexity of the analysis and exposes the physical
mechanisms of perturbation growth.  This enables us to clarify the
mechanics leading to low-angle bands in complicated problems such as
under dynamic anisotropy. We first consider the base-state, simple
shear flow at the order of $\epsilon^0$ (\S{\ref{sec:basestate}}) and
then the linearised governing equations at the order of $\epsilon^1$
(\S{\ref{sec:growthrate-solution}}). Finally, in
\S{\ref{sec:general-solution}}, we obtain the growth rate of porosity
perturbations $\phi_1$ for the most general case of dynamic
anisotropy. The result obtained is used in
\S{\ref{sec:physical-interp}} to clarify the mechanisms of
low-angle-band formation.

\subsection{Base-state simple shear flow}
\label{sec:basestate}

\begin{figure}
  \centering
  \vspace{3mm}
  \includegraphics[width=\textwidth]{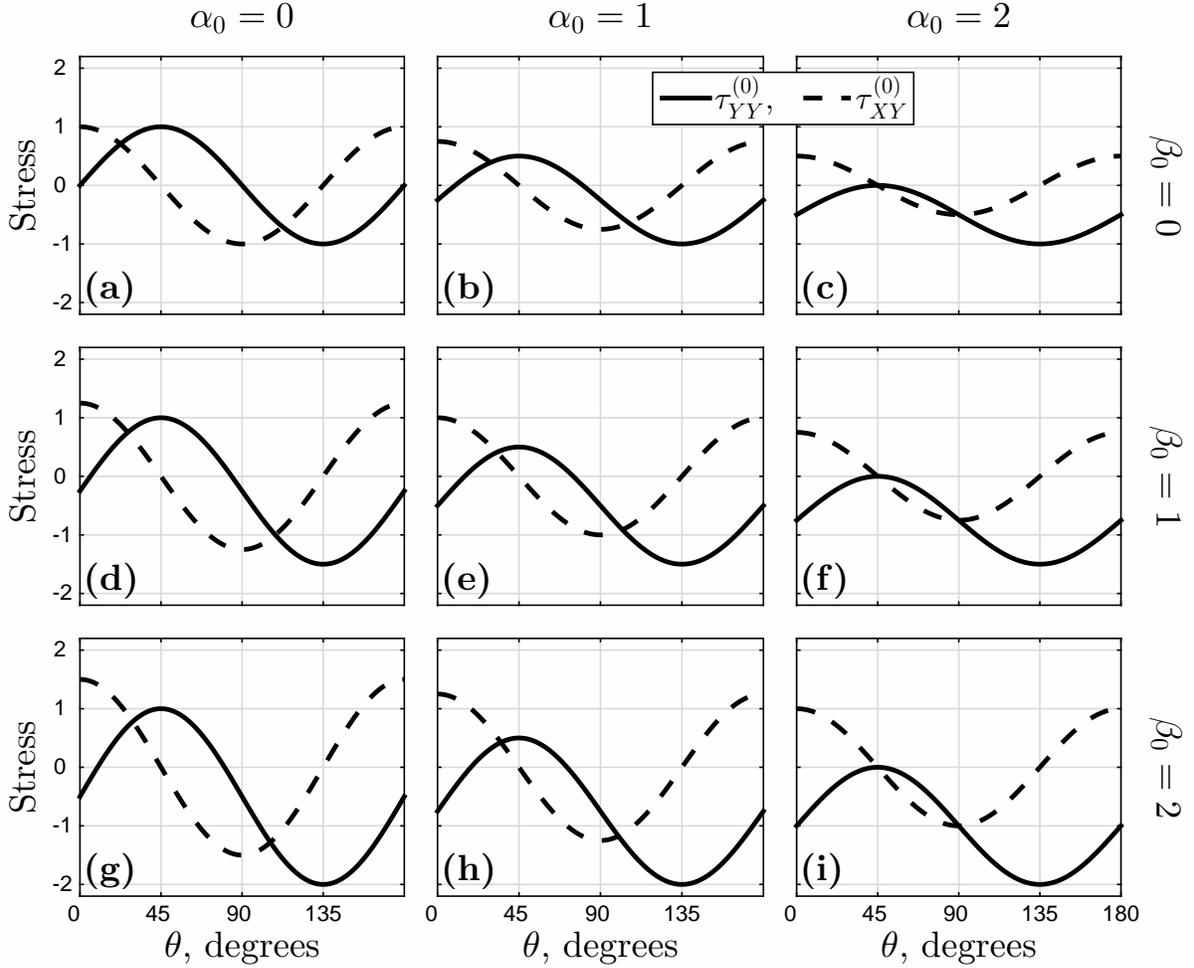}
  \caption{Normal stress $\tau^{(0)}_{YY}$ (tension positive) and
    shear stress $\tau^{(0)}_{XY}$ produced by the base-state simple
    shear flow as functions of angle $\theta$ between the shear plane
    and the $X$-axis (Fig.~\ref{fig:schematic}a).  We use
    $r_\xi=5/3$ here and throughout \citep{takei13}. Each panel is
    computed with a different set $(\alpha_0,\beta_0)$ as labelled
    above and to the right of the panels.}
  \label{fig:stress}
\end{figure}

Using the angle $\theta$ between the initial perturbation wavefronts
(aligned with the $X$ direction)
and the base-state shear plane (Fig.~\ref{fig:schematic}a), the
components of the base-state strain rate tensor in the rotated
coordinate system are
\begin{equation}
  \strr^{(0)}_{ij} = \frac{1}{2}\left(\begin{array}{cc}
     - \sin 2\theta & \cos 2\theta \\
     \cos 2\theta & \sin 2\theta
   \end{array}\right).
  \label{eq:eij0}
\end{equation}
As shown in Figs.~\ref{fig:schematic}b and \ref{fig:schematic}c,
$\tau^{(0)}_{YY}$ and $\tau^{(0)}_{XY}$ represent, respectively, the  
base-state tensile and shear stresses normal and parallel to the
perturbation wavefronts, which play important roles in understanding
the growth of these perturbations. Noting that $\strr^{(0)}_{YY} =
-\strr^{(0)}_{XX}$, these components  are given by
\begin{equation}
  \left(\begin{array}{c}
      \tau^{(0)}_{YY} \\[1mm]
      \tau^{(0)}_{XY}
    \end{array}\right) = \frac{1}{2}\left(\begin{array}{cc}
      C^{(0)}_{YYYY}-C^{(0)}_{YYXX} & 2C^{(0)}_{YYXY} \\[1mm]
      C^{(0)}_{XYYY} - C^{(0)}_{XYXX} & 2C^{(0)}_{XYXY}
    \end{array}\right)
  \left(\begin{array}{c}
      \sin2\theta \\[1mm] \cos2\theta
    \end{array}\right),
  \label{eq:bs_stress}
\end{equation}
with $\anisoangle_0 = \frac{\pi}{4}+\theta$. 
In Figure~\ref{fig:stress},  $\tau^{(0)}_{YY}$ and $\tau^{(0)}_{XY}$
are plotted as a function of angle $\theta$. An understanding of their
systematics is needed to interpret the results of the stability
analysis. 

\begin{figure}
  \centering  \vspace{3mm}
  \includegraphics[width=\textwidth]{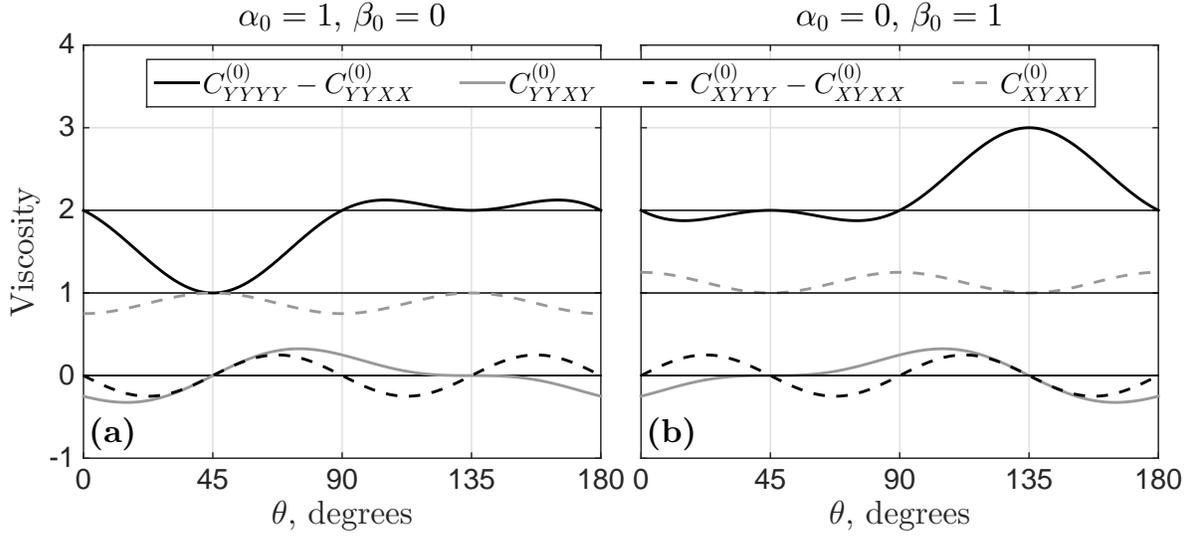}
  \caption{Viscosity components from equation~\eqref{eq:bs_stress} as
    functions of angle $\theta$ between the shear plane and the
    $X$-axis. The curves are computed for the anisotropy parameters
    \textbf{(a)} $\alpha_0=1$ and $\beta_0=0$, and \textbf{(b)}
    $\alpha_0=0$ and $\beta_0=1$. The thin black lines represent each
    component for the isotropic case ($\alpha_0=\beta_0=0$).}
  \label{fig:C}
\end{figure}

Considering first the solid curves representing normal stress, we see
that for $\alpha_0=\beta_0=0$ (isotropic, top-left panel),
$\tau^{(0)}_{YY}$ follows the expected pattern of $\sin2\theta$.  It
is tensile for $\theta<90^\circ$ and compressive for $\theta>90^\circ$
\citep{spiegelman03b}.  However, as $\alpha_0$ increases (for
$\beta_0=0$, top row of panels), $\tau^{(0)}_{YY}$ becomes negative
(compressive) at all angles.  The mechanism for this change is
two-fold. First, $\alpha_0$ decreases the viscous resistance to
extension in the $\tau_3$ direction and reduces the maximum tensile
stress. This is because increasing $\alpha_0$ reduces
$(C^{(0)}_{YYYY}-C^{(0)}_{YYXX})$ for angles near $\theta = 45^\circ$
(Figure~\ref{fig:C}a, black solid line). Superimposed on this is a
compressive stress around $\theta =0^\circ$ and $90^\circ$ that
emerges as a consequence of shear strain rate coupled to normal stress
via the $C^{(0)}_{YYXY}$ viscosity (Figure~\ref{fig:C}a, gray solid
line).  The product $C^{(0)}_{YYXY}\strr^{(0)}_{XY}$ is negative for
all $\theta$, motivating us to name this coupling
``shear-strain-induced compression.''  This non-trivial result comes
from the fact that the stress-induced softening occurs in the tensile
($\tau_3$) direction, as schematically illustrated in \cite{takei13}
Figure~7b.

The effect of increasing $\beta_0$ on $\tau^{(0)}_{YY}$ is shown down
columns in Figure~\ref{fig:stress}. Similar to $\alpha_0$, $\beta_0$
couples the shear strain rate to compressive normal stress for angles
near $\theta =0^\circ$ and $90^\circ$ via the $C^{(0)}_{YYXY}$
viscosity. In contrast to $\alpha_0$, however, $\beta_0$ strengthens
the aggregate in the $\tau_1$ direction and increases the normal
stress amplitude near $\theta = 135^\circ$ (Figures~\ref{fig:stress}
and \ref{fig:C}b). As a result, the sign-change of normal stress
caused by $\beta_0$ occurs in a limited range of
$\theta\lesssim90^\circ$ and $\theta\gtrsim0^\circ$
(Fig.~\ref{fig:stress}, bottom-left panel).

The shear stress curves in Figure~\ref{fig:stress} (dashed lines) also
change with increasing $\alpha_0$ and/or $\beta_0$.  For zero
anisotropy in panel~(a), the shear stress follows $\cos2\theta$, as
expected for coordinate rotation only. Anisotropy does not change the
mean of $\tau^{(0)}_{XY}(\theta)$, as required by symmetry of the
stress tensor.  Increasing $\alpha_0$ (or $\beta_0$) has an overall
weakening (or strengthening) effect, changing only the amplitude of
$\tau^{(0)}_{XY}$.  This is in contrast to the effect of anisotropy on
normal stress, which has a strong dependence on angle $\theta$.

\subsection{Growth of porosity perturbations}
\label{sec:growthrate-solution}

For simplicity in this linearised analysis, we consider the case of
liquid viscosity $\eta_L=0$, giving a non-dimensional compaction length
$R\rightarrow\infty$.  In this limit, liquid segregation over any
length scale occurs at vanishingly small pressure gradients.
Therefore, the pressure gradient terms in
equations~\eqref{eq:eqmo3_1st} are negligible.  Pressure
then drops out of the problem and we no longer need to solve equation
\eqref{eq:eqmo2_1st} (which has become indeterminate anyway!). This is
equivalent to considering only the subset of perturbations with
wavelengths much smaller than the dimensional compaction length
\citep[e.g.][]{katz06}.

Using the initial wavenumber vector $\boldsymbol{K}=(0,K)$, the
porosity perturbation at time $t$ is 
\begin{equation}
  \phi_1(\xvec,t)=\exp \left[i\boldsymbol{K} \cdot
    (\xvec -  {\bf V}^{(0)} t) + \dot{s}t\right],
  \label{eq:perturbation}
\end{equation}
which accounts for rotation of the wave-vector due to advection by the
base-state flow \citep{spiegelman03b}.
At $t=0$, by choice of the coordinate system, perturbations are
uniform in the $X$ direction. Therefore, partial derivatives of the
first-order quantities with respect to $X$ are zero.  Using
$\strr^{(0)}_{XX}=-\strr^{(0)}_{YY}$, $\strr^{(1)}_{XX}=0$, and
$\Grad P_1 = \boldsymbol{0}$, equations~\eqref{eq:first-order} become
\begin{subequations}
  \label{eq:first-order-simpler}
  \begin{align}
    \dot{s} &= (1-\phi_0)\strr^{(1)}_{YY}/\phi_1, \label{eq:eqmo1_1st2}\\
    0 &= \left[(C^{(1)}_{YYYY}-C^{(1)}_{YYXX})
        \strr^{(0)}_{YY}+2C^{(1)}_{YYXY} \strr^{(0)}_{XY} \right]
        +  \left[C^{(0)}_{YYYY} \strr^{(1)}_{YY}+2C^{(0)}_{YYXY}
        \strr^{(1)}_{XY}\right], \label{eq:eqmo3_1st2a}\\
    0 &= \left[(C^{(1)}_{XYYY}-C^{(1)}_{XYXX})
              \strr^{(0)}_{YY}+2C^{(1)}_{XYXY} \strr^{(0)}_{XY}\right]
        + \left[C^{(0)}_{XYYY} \strr^{(1)}_{YY}+2C^{(0)}_{XYXY}
        \strr^{(1)}_{XY}\right]. \label{eq:eqmo3_1st2b}
  \end{align}
\end{subequations}
Equations \eqref{eq:eqmo3_1st2a} and \eqref{eq:eqmo3_1st2b} are
obtained after an integration in the $Y$ direction; boundary
conditions are not needed because the domain is infinite and the
first-order fields are periodic.  

The right-hand side of equations~\eqref{eq:eqmo3_1st2a} and
\eqref{eq:eqmo3_1st2b} represent $\tau^{(1)}_{YY}$ and
$\tau^{(1)}_{XY}$, respectively. Since pressure gradients are
negligible, these stresses must be spatially uniform for the system to
be in balance. Therefore the first-order product of viscosity and
strain rate must sum to zero; viscosity reduction associated with
porosity perturbations (within the first square brackets in the right
hand side) is compensated by the strain rate perturbations (within the
second square brackets).

To facilitate the physical interpretation of
equations~\eqref{eq:eqmo3_1st2a} and \eqref{eq:eqmo3_1st2b}, these
equations are re-expressed as
\begin{equation}
  \label{eq:strategy}
  \left(
    \begin{array}{cc}
      C^{(0)}_{YYYY} & C^{(0)}_{YYXY}\\[1mm]
      C^{(0)}_{XYYY} & C^{(0)}_{XYXY}
    \end{array}
  \right)\left(
    \begin{array}{c}
      \strr^{(1)}_{YY} \\[1mm]  2\strr^{(1)}_{XY}
    \end{array}
  \right) = \left(
    \begin{array}{c}
      \tau^{(f)}_{YY} \\[1mm] \tau^{(f)}_{XY}
    \end{array} 
  \right),
\end{equation}
with equivalent (``forcing'') stresses
\begin{subequations}
  \label{eq:forcing}
  \begin{align}
    \label{eq:forcing_yy}
    \tau^{(f)}_{YY} &= -\left[(C^{(1)}_{YYYY}-C^{(1)}_{YYXX})
                    \strr^{(0)}_{YY}+2C^{(1)}_{YYXY} \strr^{(0)}_{XY} \right],
    \\
    \label{eq:forcing_xy}
    \tau^{(f)}_{XY}  &=-\left[(C^{(1)}_{XYYY}-C^{(1)}_{XYXX})
                     \strr^{(0)}_{YY}+2C^{(1)}_{XYXY} \strr^{(0)}_{XY}\right].
 \end{align}
\end{subequations}
Equation~\eqref{eq:strategy} relates the strain-rate response of the
system to the forcing stresses defined by~\eqref{eq:forcing}.  From
equation~\eqref{eq:strategy}, normal strain rate in the $Y$ direction
$\strr^{(1)}_{YY}$ (the component that is most relevant to the
perturbation growth) can be expressed as
\begin{equation}
  \label{eq:forced_strainrate}
  \strr^{(1)}_{YY} =  
  \compliance^{(0)}_{YYYY} \tau^{(f)}_{YY} + \compliance^{(0)}_{YYXY}\tau^{(f)}_{XY},
\end{equation}
where $\compliance^{(0)}_{YYYY}$ and $ \compliance^{(0)}_{YYXY}$ are
the compliances defined by
\begin{subequations}
  \label{eq:static_compliance}
  \begin{align}
    \label{eq:static_compliance_yyyy}
    \compliance^{(0)}_{YYYY} 
    & = \frac{C^{(0)}_{XYXY}}{C^{(0)}_{XYXY}C^{(0)}_{YYYY} 
      - C^{(0)}_{YYXY}C^{(0)}_{XYYY}},\\
    \compliance^{(0)}_{YYXY} 
    & = \frac{-C^{(0)}_{YYXY}}{C^{(0)}_{XYXY}C^{(0)}_{YYYY} 
      - C^{(0)}_{YYXY}C^{(0)}_{XYYY}}.
      \label{eq:static_compliance_xyyy}
  \end{align}
\end{subequations}
The forcing stresses, $\tau^{(f)}_{YY}$ and $\tau^{(f)}_{XY}$, are not
externally applied (like those causing simple shear), nor are they the
first-order stress perturbations $\tau^{(1)}_{YY}$ and
$\tau^{(1)}_{XY}$ (these are both equal to zero).  Instead, they are
equivalent stresses that are created internally as a consequence of
the base-state flow acting on the viscosity change associated with
porosity perturbations. Moreover, under dynamic anisotropy, these
forcing terms also depend on the strain-rate perturbations and hence
equation (\ref{eq:forced_strainrate}) does not always give an explicit
solution for $ \strr^{(1)}_{YY} $. Nonetheless,
equation~\eqref{eq:forced_strainrate} enables us to 
separate the mechanics into two simpler parts: the forcing,
$\tau^{(f)}_{YY}$ and $\tau^{(f)}_{XY}$, and the compliance,
$\compliance^{(0)}_{YYYY} $ and $\compliance^{(0)}_{YYXY}$, where the
latter represents the system response to forcing with unit amplitude.
This decomposition is helpful to understand the detailed (and rather
complicated) mechanisms of the different models considered here.

\subsection{General solution}
\label{sec:general-solution}

Equations~\eqref{eq:first-order-simpler} are solved here to obtain
an explicit expression for $\dot{s}$ for the full model of dynamic
anisotropy. The first-order viscosity tensor
is written in terms of the porosity and anisotropy perturbations
$\phi_1$, $\alpha_1$, $\beta_1$, and $\Theta_1$. The anisotropy
perturbations are then expressed in terms of the porosity and
strain-rate perturbations $\phi_1$, $\strr^{(1)}_{YY}$, and
$\strr^{(1)}_{XY}$. These calculations are sketched in
Appendix~\ref{sec:calculation-details}. The components of the
first-order viscosity tensor are then substituted into
equations~\eqref{eq:eqmo3_1st2a} and \eqref{eq:eqmo3_1st2b}, which are
manipulated to solve for $\strr^{(1)}_{YY}$ and $\strr^{(1)}_{XY}$ as
functions of $\phi_1$.  The normal strain rate $\strr^{(1)}_{YY}$
obtained by this approach is substituted into \eqref{eq:eqmo1_1st2} to
give an expression for the growth-rate of perturbations,
\begin{equation}
  \dot{s}  = (1-\phi_0)\lambda\left[\compliance^{(0)}_{YYYY}\dyfac_{p}
    \left(\tau_{YY}^{(0)} -q\tau_{XX}^{(0)}\right)
    + \compliance^{(0)}_{YYXY}\dyfac_{q}\left(\tau_{XY}^{(0)}
      -p\tau_{XX}^{(0)}\right)\right],
  \label{eq:growthrate_dynamicaniso}
\end{equation}
with the compliances defined by   \eqref{eq:static_compliance} and
dynamic factors 
\begin{subequations}
  \begin{align}
    \label{eq:dynamic_factors}
    \dyfac_p &= \left(1-p{C^{(0)}_{XXXY}}/{C^{(0)}_{XYXY}}\right)/{D},\\
    \dyfac_q &= \left(1-q{C^{(0)}_{XXXY}}/{C^{(0)}_{YYXY}}\right)/{D},
  \end{align}
\end{subequations}
where 
\begin{equation}
  \label{eq:determinant}
  D = 1 + q\left(\tfrac
    {C^{(0)}_{XXXY}C^{(0)}_{XYYY} - C^{(0)}_{XYXY}C^{(0)}_{XXYY}}
    {C^{(0)}_{XYXY}C^{(0)}_{YYYY} - C^{(0)}_{YYXY}C^{(0)}_{XYYY}}\right) 
  + p\left(\tfrac
    {C^{(0)}_{YYXY}C^{(0)}_{XXYY} - C^{(0)}_{XXXY}C^{(0)}_{YYYY}}
    {C^{(0)}_{XYXY}C^{(0)}_{YYYY} - C^{(0)}_{YYXY}C^{(0)}_{XYYY}}\right).
\end{equation}
The constant coefficients $p$ and $q$ that express the sensitivity of
the growth rate to dynamic anisotropy perturbations are
\begin{subequations}
  \label{eq:constants}
  \begin{align}
    p&=\zeta\displaystyle\frac{1-r_\beta}{4}\left( 2\cos 2\theta-
       \frac{\alpha_0\Delta\tau^{(0)}}{ \daddtau
       (\tau_{XX}^{(0)}-\tau_{YY}^{(0)})^2}\sin 2\theta \sin 4\theta\right),\\
    q&=\zeta\frac{1-r_\beta}{4}\left(2\sin2\theta
       + \frac{\alpha_0\Delta\tau^{(0)}}{ \daddtau
       (\tau_{XX}^{(0)}-\tau_{YY}^{(0)})^2}\cos 2\theta \sin
       4\theta\right) + \zeta\frac{1+r_\beta}{2},
  \end{align}
\end{subequations}
where $\daddtau$ and $\zeta$ are defined by equations
(\ref{eq:alphacap}) and (\ref{eq:zeta}), respectively. We do not
attempt to physically interpret the detailed form of $p$ and $q$. It
is important to note, however, that for the static anisotropy model,
$p$ and $q$ are zero, and $\dyfac_p=\dyfac_q=1$.

\section{Physical interpretation in various limits}
\label{sec:physical-interp}

The growth rate in eqn.~\eqref{eq:growthrate_dynamicaniso} is a
general result for the full model presented in
\S{\ref{sec:governing-eqns}} above (with the sole assumption of
$R\to\infty$).  To build up a physical understanding of this equation,
we return to the simpler case of static anisotropy, which includes the
simplest case of the Newtonian, isotropic model.  The static
anisotropy model has previously been studied by \cite{takei13} using
linearised analysis. However, the mathematical complexity of their
results precluded a detailed mechanical interpretation. A
reconsideration using the perturbation-oriented coordinate system
enables a physical understanding of the instability mechanism and the
rheological control on the dominant band angle.  These are needed to
understand the more complicated, dynamic model. To facilitate this (in
\S{\ref{sec:static}--\ref{sec:dynamic}}), we make the simplifying
assumption that $\beta=0$ --- that there is no contiguity increase in
the $\tau_1$-direction. The effect of non-zero $\beta$ is discussed in
section~\ref{sec:beta}, where we show that its role is minor compared
to that of $\alpha$, the contiguity decrease in the
$\tau_3$-direction.

\subsection{Static anisotropy}
\label{sec:static}

When $\alpha_1=\beta_1=\anisoangle_1=0$, the mechanical equilibrium
conditions \eqref{eq:eqmo3_1st2a} and \eqref{eq:eqmo3_1st2b} are
written as
\begin{subequations}
  \label{eq:static2}
  \begin{align}
    C^{(0)}_{YYYY} \strr^{(1)}_{YY} + 2C^{(0)}_{YYXY}\strr^{(1)}_{XY} 
    &=  \lambda \phi_1 \tau^{(0)}_{YY}  \\
    C^{(0)}_{XYYY} \strr^{(1)}_{YY}  +2C^{(0)}_{XYXY}\strr^{(1)}_{XY}
    &= \lambda \phi_1 \tau^{(0)}_{XY} .
  \end{align}
\end{subequations}
Comparison with equation~\eqref{eq:strategy} shows that the forcing
stresses are given by
$\tau^{(f)}_{YY} = \lambda \phi_1\tau^{(0)}_{YY}$ and
$\tau^{(f)}_{XY} = \lambda \phi_1\tau^{(0)}_{XY}$. These forcing
stresses are caused by the base-state tensile and shear stresses
acting on the porosity perturbation by way of porosity weakening
rheology ($\lambda>0$) as depicted in Figs.~\ref{fig:schematic}b,c. In
this simple model, $\tau^{(f)}_{YY} $ and $\tau^{(f)}_{XY} $ are given
in terms of the porosity perturbation $\phi_1$, and hence
eqn.~\eqref{eq:forced_strainrate} provides an explicit solution for
$\strr^{(1)}_{YY}$.

It is evident from \eqref{eq:eqmo1_1st2} that the normal strain rate
$\strr^{(1)}_{YY}$ causes an increase in the amplitude of porosity
perturbations; the shear strain rate $\strr^{(1)}_{XY}$ does not cause
the porosity to change. When the viscosity is anisotropic, both the
normal and the shear stress drive $\strr^{(1)}_{YY}$ and hence
contribute to perturbation growth. This does not occur under isotropic
viscosity.

\subsubsection{Instability mechanism in the isotropic system}
\label{sec:isotropy}

For an isotropic aggregate ($\alpha_0=\beta_0=0$),
$C^{(0)}_{YYXY}=C^{(0)}_{XYYY}=0$ in equations~\eqref{eq:strategy} and
\eqref{eq:static2}, and the compliance $\compliance^{(0)}_{YYXY}$ that
couples shear stress to normal strain rate is zero.  In this case,
$\strr^{(1)}_{YY}$ is driven only by the base-state normal stress
$\tau_{YY}^{(0)}$.  The growth rate $\dot{s}$ is given by
\begin{equation}
  \dot{s}  = (1-\phi_0) \lambda\frac{\tau_{YY}^{(0)}}{C_{YYYY}^{(0)}} 
  \hspace{10mm} \mbox{isotropic model.}
  \label{eq:growthrate_iso}
\end{equation}
Perturbations are unstable under tensile stress ($\tau_{YY}^{(0)}>0$)
normal to the perturbation wavefronts, stable under compressive normal
stress ($\tau_{YY}^{(0)}<0$), and unaffected by shear stress
$\tau_{XY}^{(0)}$.  We therefore term this the tensile-stress-induced
instability, or tensile instability.  When band angle $\theta$
relative to the simple shear flow is $45^\circ$, the tensile stress
$\tau_{YY}^{(0)}$ attains its maximum (Figure~\ref{fig:stress}a) and
hence the growth rate $\dot{s}$ is also at a maximum, as shown in
Figure~\ref{fig:static}a and by \cite{spiegelman03b}. The occurrence
of the tensile instability in a porosity-weakening, two-phase
aggregate was first predicted by \cite{stevenson89}.

\subsubsection{Two instability mechanisms in the anisotropic system}

\begin{figure}
  \centering \vspace{3mm}
  \includegraphics[width=\textwidth]{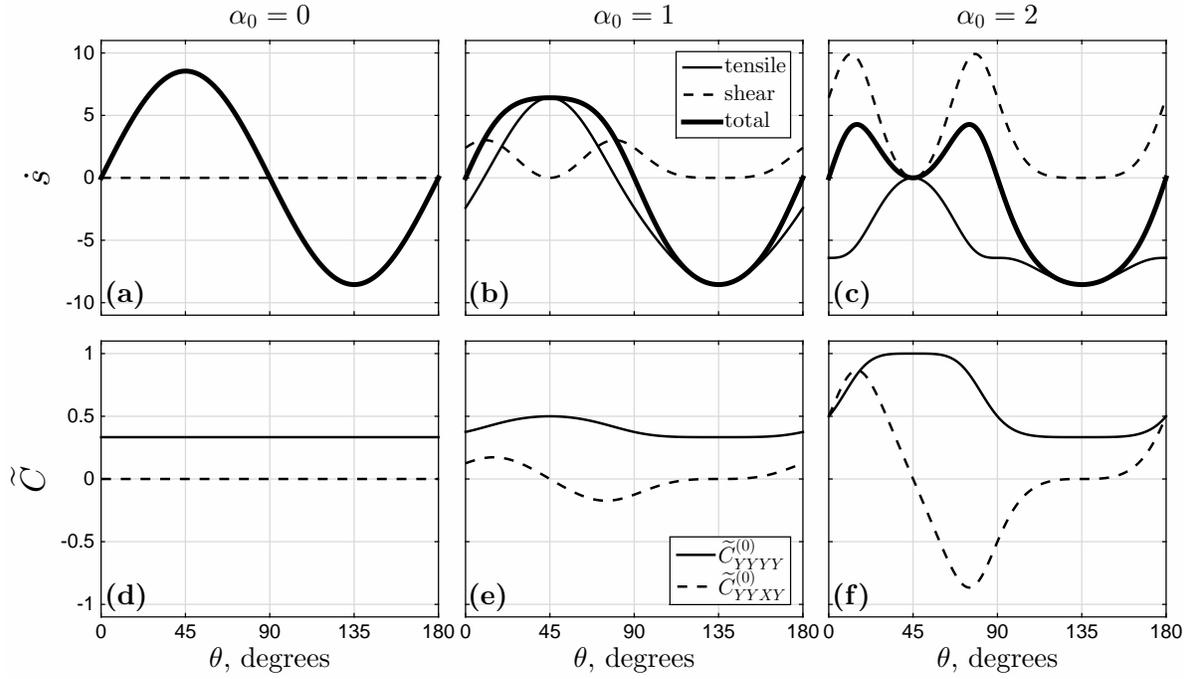}
  \caption{Characteristics of the static anisotropy model as a
    function of the angle between porosity perturbations and the shear
    plane. Each column is for a different value of $\alpha_0$.  In all
    cases, $\beta_0=0$, $\lambda=27$, and $\phi_0=0.05$. \textit{Top
      row.} Growth rate $\dot{s}$ of perturbations $\phi_1$ from
    eqn.~\eqref{eq:growthrate_aniso}. The heavy line represents the
    total growth rate; light lines show the growth rate decomposed
    into two parts: the contribution from the tensile instability (1st
    term of eqn.~\eqref{eq:growthrate_aniso}) and that from the shear
    instability (2nd term). \textit{Bottom row.}  Compliances
    $\compliance^{(0)}_{YYYY}$ and $\compliance^{(0)}_{YYXY}$ in
    eqn.~\eqref{eq:growthrate_aniso}.}
 \label{fig:static}
\end{figure}

For an anisotropic aggregate ($\alpha_0 > 0$ and/or $\beta_0 > 0$),
there is a coupling between shear and normal components via
$C^{(0)}_{YYXY}=C^{(0)}_{XYYY}\neq 0$. In this case,
$\strr^{(1)}_{YY}$ is forced by both normal stress across
perturbations and shear stress along perturbations. The growth rate
is
\begin{equation}
  \dot{s}  = (1-\phi_0) \lambda \left(
  \compliance^{(0)}_{YYYY}\tau_{YY}^{(0)} +
  \compliance^{(0)}_{YYXY}\tau_{XY}^{(0)}\right) 
  \hspace{10mm} \mbox{static anisotropy model,}
  \label{eq:growthrate_aniso}
\end{equation}
using the compliances given by equations~\eqref{eq:static_compliance}.
The first term on the right-hand side of \eqref{eq:growthrate_aniso}
represents the tensile instability, generalised to the anisotropic
aggregate. The 2nd term represents a shear-stress-induced instability
that does not occur in the isotropic system. The total growth rate
$\dot{s}$ versus band angle $\theta$ is plotted in the top row of
panels of Figure~\ref{fig:static} for various anisotropy amplitudes
$\alpha_0$ (thick lines).  Consistent with previous work, as
$\alpha_0$ is increased, the single growth rate peak splits into two
peaks at low and high angles to the shear plane
(Fig.~\ref{fig:static}c).  Because the lower-angle peak dominates the
higher angle peak after a finite time \citep{katz06, takei13}, this
result means a significant lowering of the dominant band angle by the
viscous anisotropy --- if the magnitude of anisotropy $\alpha$ is
sufficiently close to saturation ($\alpha \simeq 2$).

\subsubsection{How viscous anisotropy causes lowering of band angle}

Although the effect of viscous anisotropy is evident from the total
growth rate shown in the top panels of Figure~\ref{fig:static}, it is
not immediately obvious why the dominant band angle is lowered by
viscous anisotropy. The physical mechanism can be understood by
considering the tensile and shear components of the growth rate
independently (1st and 2nd terms of \eqref{eq:growthrate_aniso},
respectively). In the top row of Figure~\ref{fig:static}, these two
growth rates are plotted separately for various values of $\alpha_0$
(thin solid curve for tensile instability; thin dashed curve for shear
instability). Comparison of panels (a), (b), and (c) reveals that the
peak split occurs through (\textit{i}) stabilisation of the tensile
instability and (\textit{ii}) emergence of the shear instability with
increasing magnitude of anisotropy~$\alpha$. We consider each of these
in turn.

To understand why viscous anisotropy stabilises the tensile
instability, we return to the systematics of the base-state stress
(section \ref{sec:basestate}). Comparison of the three columns of
Fig.~\ref{fig:stress} shows that as $\alpha_0$ increases, the tensile
stress $\tau^{(0)}_{YY}$ decreases in amplitude and becomes
compressive at all angles.  With $\tau^{(0)}_{YY}\le 0$, the first
term in equation~\eqref{eq:growthrate_aniso} is always less than or
equal to zero, and hence stable.

To understand why viscous anisotropy destabilises the shear mechanism,
we consider the coupling between the shear stress that drives the
instability and the normal strain-rate that is responsible for its
growth.  As shown by equation~\eqref{eq:forced_strainrate} with
$\tau^{(f)}_{YY} = \lambda \phi_1\tau^{(0)}_{YY}$ and
$\tau^{(f)}_{XY} = \lambda \phi_1\tau^{(0)}_{XY}$, the shear stress
$\tau^{(0)}_{XY}$ is coupled to the normal strain rate
$\strr^{(1)}_{YY}$ via $\compliance^{(0)}_{YYXY}$. The angular
dependence of $\compliance^{(0)}_{YYXY}$ is shown by dashed curves in
the bottom-row panels of Figure~\ref{fig:static}.  If
$\compliance^{(0)}_{YYXY}\tau^{(0)}_{XY}$ is positive, then $\dot{s}$
is positive (or $\strr^{(1)}_{YY}$ is in phase with $\phi_1$) and the
shear mechanism contributes to unstable growth of porosity
perturbations. In fact, this product is positive (or zero) for all
$\theta$, enabling us to name this coupling ``shear stress-induced
expansion.'' This non-trivial result comes from the assumed
microstructural behaviour: that stress-induced softening occurs in the
tensile ($\tau_3$) direction, as illustrated in \cite{takei13}
Figure~7a. So the porosity perturbation grows because of the shear
mechanism, for which the low angle is favorable.

\subsubsection{Summary of static anisotropy model} 

As a recap and summary, note that under isotropic viscosity, the
growth of bands at $45^\circ$ to the shear plane is caused by a
tensile instability \citep{stevenson89, spiegelman03b}.  In contrast,
under anisotropic viscosity, the peak growth rate of bands is
controlled by a distinct shear instability. Although the peak growth
rate of the shear instability occurs at $\theta<15^{\circ}$,
stabilisation at these low angles by the tensile mechanism acts to
give a maximum in the combined growth rate at $\theta=15^{\circ}$.

The comparison between isotropic and anisotropic systems developed
above is summarised in the first two rows of
Table~\ref{table:summary}. The tensile instability is separated into
porosity-weakening $\lambda$, which is fundamental to all models, and
the tensile stress across bands $\tau_{YY}^{(0)}$, which affects both
isotropic and anisotropic cases. A shift of $\tau_{YY}^{(0)}$ to more
negative, compressive values (represented by $\bigtriangleup$)
stabilises the tensile instability. In contrast, the difference in
shear instability can be simply shown by its existence or
non-existence ($\bigcirc$~or~--).  It is the leading-order terms
$\tau_{YY}^{(0)}$ and $\compliance^{(0)}_{YYXY}$ that are responsible
for these differences.
 
\cite{katz06} extended the analysis of isotropic viscosity to include
a power-law dependence of viscosity on strain rate (or, equivalently,
on stress). They showed that strain-rate weakening viscosity leads to
lowering of band angle. In the discussion section, we compare the
angle-lowering mechanism of viscous anisotropy to that of the
power-law viscosity.  This is enabled by a reanalysis of the power-law
model using the rotated coordinate system.

\begin{table}
  \small
  \caption{Summary of band formation models. $\bigcirc$: exists; --:
    does not exist; $\bigtriangleup$: is modified.}
  \begin{tabularx}{\textwidth}{@{} lccYcc @{}} 
    \hline
    &\multicolumn{3}{c}{$\compliance^{(0)}_{YYYY}\,\tau^{(f)}_{YY}$ (tensile)} 
    & $\compliance^{(0)}_{YYXY}\,\tau^{(f)}_{XY}$ (shear) &   \\ \addlinespace
    \cline{2-4} Model  & $\lambda>0$ &  $\tau^{(0)}_{YY} >0$ 
                & additional factor    &   &  Dominant angle \\
    \hline   
    Isotropic Newtonian &$\bigcirc$&$\bigcirc$&--&--&$45^{\circ}$ \\
    Static anisotropy
    &$\bigcirc$&$\bigtriangleup$&--&$\bigcirc$ &$\sim 23^{\circ}$ \\
    Dynamic anisotropy
    &$\bigcirc$&$\bigtriangleup$&$\bigcirc$&$\bigcirc$ &$\sim 10^{\circ}$\\
    Isotropic power-law 
    &$\bigcirc$&$\bigcirc$&$\bigcirc$&--&$\sim 20^{\circ}$  
  \end{tabularx}
  \label{table:summary}
\end{table}

\subsection{Dynamic anisotropy}
\label{sec:dynamic}

We return to the full expression for the growth rate of bands,
eqn.~\eqref{eq:growthrate_dynamicaniso}, to develop a physical
understanding of why dynamic anisotropy lowers band angles, as
observed in the numerical solutions (Fig.~\ref{fig:numerical}).  To do
so we take $\alpha_0=1$ and again make the simplifying assumption that
$\beta=0$ (though see \S{\ref{sec:beta}}, below).

The perturbations in $\Theta$ and $\alpha$ under dynamic anisotropy
are obtained by linearisation of equations~\eqref{eq:Theta} and
\eqref{eq:alphabeta} with respect to the stress perturbation
$\tau_{ij}^{(1)}$. The expansion is conducted around the base-state
values $\Theta_0,\alpha_0$.  In
Appendix~\ref{sec:calculation-details}, we show that the sensitivity
of $\alpha$ to variations in deviatoric stress is given by the
parameter
\begin{equation}
  \label{eq:alpha_sensitivity}
  \daddtau =
  \left.\pdiff{\alpha}{\Delta\tau}\right\vert_{\Delta\tau^{(0)}}
  = \frac{2}{\tausat}\text{sech}^2\left(\frac{2(\Delta\tau^{(0)}-\tauoff)}
    {\tausat}\right).
\end{equation}
This parameter allows us to write
$\alpha_1 = \daddtau\Delta\tau^{(1)}$, and hence to see that static
anisotropy corresponds to the case where $\daddtau = 0$.  The
situation for $\Theta$ is more complicated because there is no single
parameter that controls its sensitivity to deviatoric stress;
variations of $\Theta$ can either be fully considered or fully
neglected.  Fortunately, numerical and analytical results show that
these variations ($\Theta_1$) play an insignificant role in the
understanding of band angles, and hence we need consider only the
magnitude of anisotropy $\alpha$.  This is achieved by looking at the
dependence of key quantities (especially $\dot{s}$) on $\daddtau$.

\begin{figure}
  \centering \vspace{3mm}
  \includegraphics[width=\textwidth]{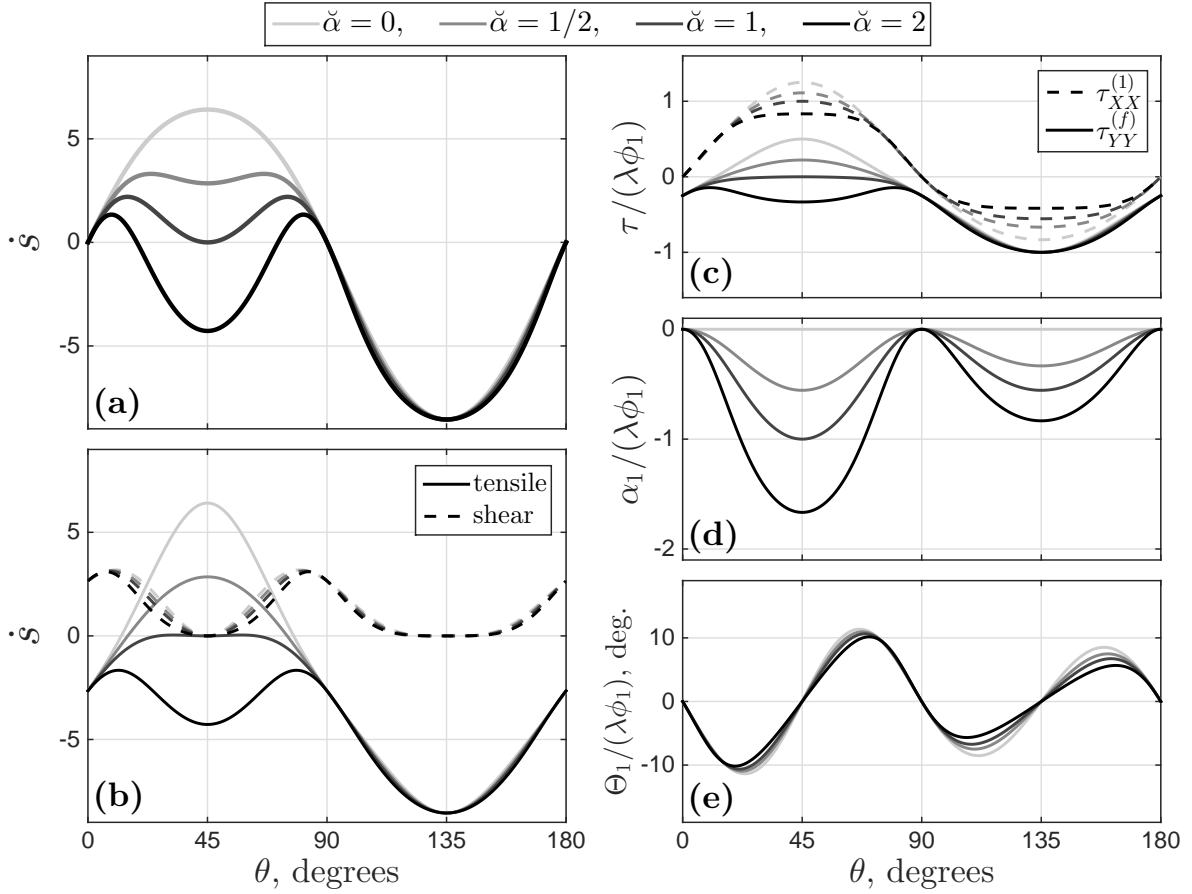}
  \caption{Characteristics of the dynamic anisotropy model for various
    values of $\daddtau$ as a function of the angle $\theta$ between
    porosity perturbations and the shear plane.  In all panels,
    $\alpha_0=1$, $\beta_0=0$, $\lambda=27$, and $\phi_0=0.05$.
    \textbf{(a)} Full growth-rate $\dot{s}$ from
    eqn.~\eqref{eq:growthrate_dynamicaniso}. \textbf{(b)} Growth-rate
    from eqn.~\eqref{eq:growthrate_dynamicaniso} split into the
    tensile-instability term (solid lines) and the shear-instability
    term (dashed lines). \textbf{(c)} Band-normal forcing stress
    (solid lines; eqn.~\eqref{eq:forcing_yy}) and band-parallel
    perturbation stress (dashed lines; eqn.~\eqref{eq:stressXX}).
    Both are divided by $\lambda\phi_1$. Calculation details are in
    Appendix~\ref{sec:calculation-details}. \textbf{(d)} and
    \textbf{(e)} The anisotropy perturbations $\alpha_1$ and
    $\anisoangle_1$ divided by $\lambda\phi_1$, calculated according
    to eqn.~\eqref{eq:alpha1_Theta1_II}.} 
  \label{fig:dynamic}
\end{figure}

The growth-rate of porosity perturbations $\dot{s}$ is shown in
panel~(a) of Figure~\ref{fig:dynamic} for $\alpha_0=1$ and for values
of $\daddtau$ ranging from zero to two.  Although $\daddtau=0$ does
not exclude linearised variations in $\Theta$, comparison with the
$\dot{s}$-curve in Fig.~\ref{fig:static}b confirms that variations in
$\Theta$ are ineffectual; with $\daddtau=0$ the expected band angle is
$45^\circ$.  For increasing $\daddtau$, the growth-rate peak again
splits into peaks at low and high angles.  It is important to note
that the mean value of $\alpha$ ($\alpha_0$) is not changed in this
exercise --- only the amplitude of variations about that mean.
Consistent with the numerical results of Figure~\ref{fig:numerical},
dynamic variations in the magnitude of anisotropy can sharply reduce
band angles, even at moderate $\alpha_0$ for which the static
anisotropy model predicts a high band angle ($45^\circ$).

Figure~\ref{fig:dynamic}b breaks the full growth rate into two parts,
each associated with one of the terms of
equation~\eqref{eq:growthrate_dynamicaniso}. Dashed lines,
representing the shear instability, are almost unaffected by
$\daddtau$. In contrast, the tensile instability is strongly
stabilised with increasing $\daddtau$.  This stabilisation causes the
peak of the full growth rate in panel~(a) to split into low- and
high-angle peaks.  To understand why dynamic anisotropy promotes low
band angles, it is therefore sufficient to understand why it
stabilises the tensile instability.

The tensile instability is driven by $\tau^{(f)}_{YY}$, as discussed
in \S{\ref{sec:static}}.  This represents the normal stress (tension
positive) that arises when viscosity perturbations interact with the
base-state strain rate. The detailed form of the forcing stress for
the dynamic anisotropy model is given in
equation~\eqref{eq:balance_dynamic}.  Figure~\ref{fig:dynamic}c shows
that the forcing normal stress $\tau^{(f)}_{YY}$ varies significantly
with $\daddtau$ (whereas the forcing shear stress, not shown, is
almost unaffected by $\daddtau$).  The system compliances, which are
leading-order quantities, are not affected by dynamic
anisotropy. Therefore, it is the variation of $\tau^{(f)}_{YY}$ that
is responsible for stabilisation of the tensile instability under
dynamic anisotropy.

To develop a physical understanding of the detailed dependence of
$\tau^{(f)}_{YY}$ on $\phi_1$, $\alpha_1$ and $\anisoangle_1$
(eqn.~\eqref{eq:balance_dynamic_yyyy}), focus attention on
$\theta=45^\circ$, as this is the dominant band angle when
$\daddtau=0$.  For bands at $45^\circ$, solid curves in panel~(c) of
Figure~\ref{fig:dynamic} show that the forcing stress goes from a
positive perturbation (in phase with $\phi_1$) to a negative
perturbation (anti-phased with $\phi_1$) with increasing $\daddtau$
--- hence the forcing stress $\tau_{YY}^{(f)}$ in the high-porosity
bands goes from tensile to compressive.  This change is due to an
increase in the magnitude of anisotropy perturbation
$\alpha_1 = \daddtau\Delta\tau^{(1)}$, shown in panel~(d).  Since
$\tau^{(1)}_{XY} = \tau^{(1)}_{YY} = 0$, the deviatoric stress
perturbation $\Delta\tau^{(1)}$ is entirely due to the band-parallel
normal stress perturbation $\tau^{(1)}_{XX}$ (according to
eqn.~\eqref{eq:stressXX}), which is shown by dashed curves in
panel~(c). Because $\tau^{(0)}_{XX}<0$, $\tau^{(1)}_{XX}>0$ signifies
a magnitude reduction of $\tau_{XX}$ in the high-porosity bands; the
largest change occurs for bands at $\theta=45^{\circ}$. As sensitivity
$\daddtau$ to deviatoric stress increases, $\alpha_1$ becomes more
negative (panel~(d)).  Negative values of $\alpha_1$ (anti-phased with
$\phi_1$) mean high-porosity bands have lower deviatoric stress and
weaker anisotropy than the low-porosity, inter-band regions. This is
consistent with numerical results in Fig.~\ref{fig:numerical}d.

Figure~\ref{fig:static}d--f shows that $\alpha_0$ increases the normal
compliance $\compliance^{(0)}_{YYYY}$ at angles between zero and
90$^\circ$. A negative perturbation to $\alpha_0$ therefore makes the
high-porosity bands in this range of angles less compliant to tensile
stress and the low-porosity inter-bands more compliant. Overall, then,
the perturbation in anisotropy amplitude $\alpha_1$ tends to cancel
the direct effect of the porosity perturbation $\phi_1$ on the normal
compliance, and hence $\alpha_1$ works to stabilise the tensile
instability.

The comparison between the static and dynamic anisotropy models
developed in this section is summarised in
Table~\ref{table:summary}. These two models are identical at leading
order but different at first order. Therefore, stabilisation of the
tensile mechanism due to more compressive base-state stress
($\tau_{YY}^{(0)}$) and destabilisation of the shear mechanism due to
shear stress-induced expansion ($\compliance^{(0)}_{YYXY}$) occur in
both the static and the dynamic anisotropy model. These two cases
differ, however, in that further stabilisation of the tensile
mechanism occurs due to the dynamic variation of anisotropy magnitude
($\alpha_1$). This effect hardens the band regions and weakens the
inter-band regions under dynamic anisotropy. This additional factor
($\bigcirc$ in Table~\ref{table:summary}) significantly lowers the
band angle.

It is interesting to note that dynamic perturbations to the angle of
anisotropy $\anisoangle_1$ are not an important control on band
angle. Figure~\ref{fig:dynamic}e shows that they are not affected by
$\daddtau$.  More importantly, $\anisoangle_1$ is always zero for
bands orientated at $\theta=45^\circ$.  This indicates that the
stabilisation of the tensile instability and the lowering of band
angle under dynamic anisotropy cannot be attributed to
$\anisoangle_1$.  In numerical simulations
(Fig.~\ref{fig:numerical}e), the variations of $\anisoangle$ do not
contribute to the lowering of band angle that is observed in
Fig~\ref{fig:numerical}c, though they are well-explained by the
stability analysis at $\theta\simeq 10^\circ$ (red dashed line).

\subsection{The effect of contiguity increase in the
  $\tau_1$-direction} 
\label{sec:beta}

Until now, we have neglected $\beta$ and focused on the effects of
$\alpha$, which quantifies contiguity decrease in the direction of
maximum tension. Non-zero $\alpha$ represents a weakening in the
$\tau_3$ direction that (\textit{i}) reduces the magnitude of tensile
stress and leads to (\textit{ii}) shear strain-induced compression and
(\textit{iii}) shear stress-induced extension \citep{takei13}. We have
shown that the tensile mechanism is stabilised around
$\theta=45^\circ$ by the first of these and is stabilised around
$\theta=0^\circ$ and $90^\circ$ by the second; we have also shown that
the shear mechanism is destabilised by (\textit{iii}). The parameter
$\beta$ quantifies the contiguity increase in the direction of maximum
compression. Even if $\alpha$ is zero, a non-zero $\beta$ creates
viscous anisotropy (see eqn.~\eqref{eq:Csystem}), causing the
couplings (\textit{ii}) and (\textit{iii}).  However, Fig.~\ref{fig:C}
shows that $\beta_0$ does not cause the weakening (\textit{i}). It is
this weakening, by $\alpha$ only, that is responsible for splitting
the growth-rate peak in both static and dynamic models. On this basis,
we expect the effect of $\beta_0$ to be small. This is indeed the
case: as shown below, $\beta$ alone does not cause a lowering of band
angle, but it can affect the lowering by $\alpha$.

\begin{figure}
  \centering  \vspace{3mm}
  \includegraphics[width=\textwidth]{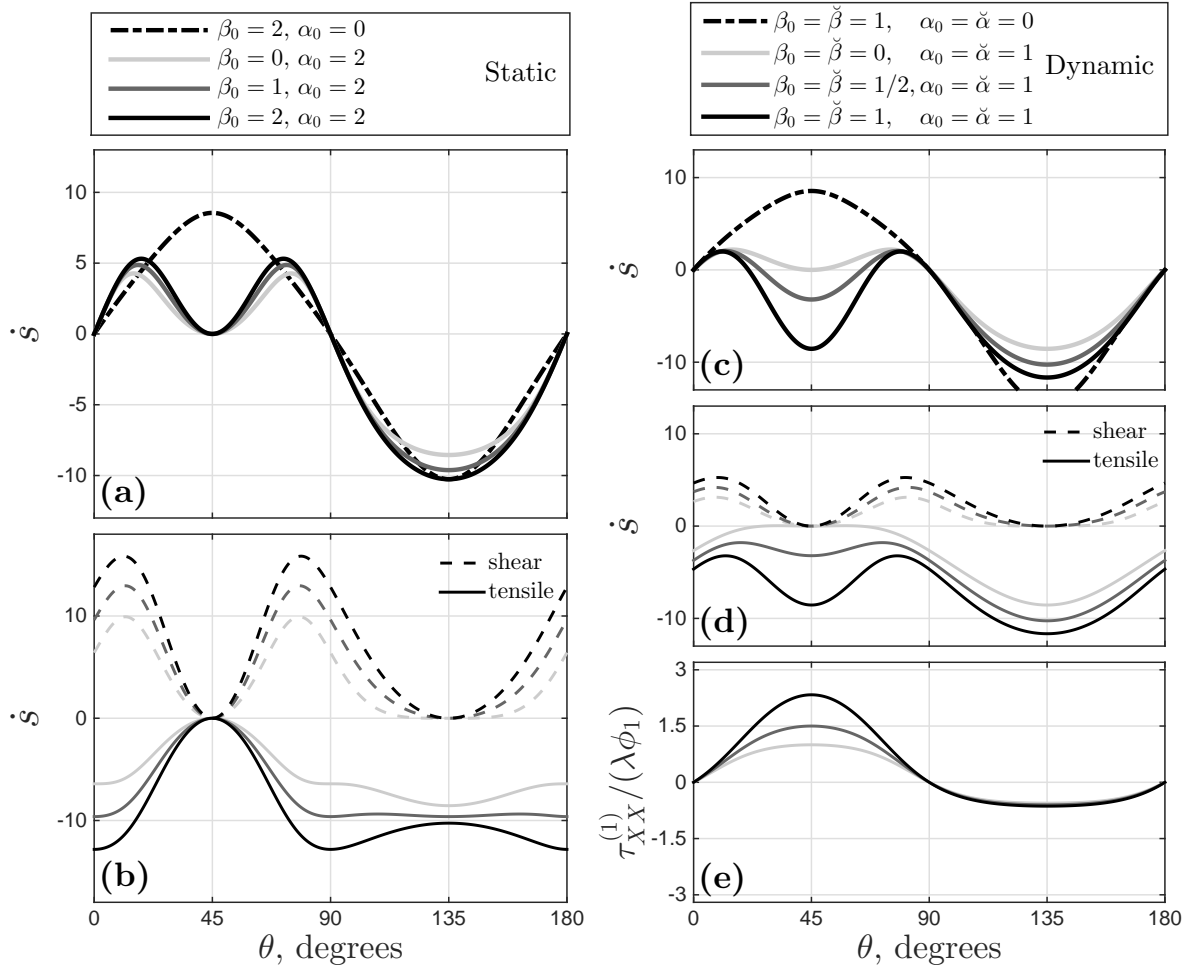}
  \caption{The effect of $\beta$ on the growth rate of porosity
    perturbations for static (left column) and dynamic (right column)
    anisotropy. \textbf{(a)}~Full growth rate $\dot{s}$ from
    eqn.~\eqref{eq:growthrate_aniso} for $\alpha_0=2$ and various
    values of $\beta_0$. A single curve for $\alpha_0=0$ and
    $\beta_0=2$ is also shown. In panel~\textbf{(b)}, the cases with
    $\alpha_0=2$ are decomposed into tensile and shear parts. Line
    greyscale has the same meaning as in panel~(a); there is no curve
    for $\alpha_0=0$. \textbf{(c)}~Full growth rate $\dot{s}$ from
    eqn.~\eqref{eq:growthrate_dynamicaniso} for $\alpha_0=\daddtau=1$
    and various values of $\beta_0=\dbddtau$.  A single curve for
    $\alpha_0=\daddtau=0$ and $\beta_0=\dbddtau=1$ is also plotted.
    In panel~\textbf{(d)}, cases with $\alpha_0=\daddtau=1$ are
    decomposed into tensile and shear parts. \textbf{(e)} The
    band-parallel, normal stress perturbation $\tau^{(1)}_{XX}$.}
  \label{fig:beta}
\end{figure}

The left column of Figure~\ref{fig:beta} illustrates the effect of
$\beta$ under static anisotropy. Panel~(a) shows that under static
anisotropy, $\dot{s}$ is split into high and low angle peaks for any
value of $\beta_0=0$ when $\alpha_0=2$ (solid curves), whereas it is
peaked at $45^\circ$ for any value of $\beta_0$ when $\alpha_0=0$
($\beta_0=2$ shown by dash-dot curve).  For $\alpha_0=2$, increasing
$\beta_0$ causes a modest shift to more compressive $\tau_{YY}^{(0)}$
at $\theta\sim0^\circ$ and $\sim90^\circ$ and a modest increase in the
amplitude of shear stress $\tau_{XY}^{(0)}$ (Fig.~\ref{fig:stress},
right column). Therefore, as Figure~\ref{fig:beta}b shows, $\beta_0$
causes stabilisation of the tensile instability and destabilisation of
shear instability in equal measure. These two effects compensate each
other and the solid growth-rate curves in Figure~\ref{fig:beta}a are
thus all very similar to that for $\beta_0=0$.

The right column of Figure~\ref{fig:beta} shows how $\beta$ affects
dynamic anisotropy. Panel~(c) shows that for $\dot{s}$ in the dynamic
anisotropy model, a two-peaked growth rate occurs for
$\alpha_0=\daddtau=1$ and $\beta_0=\dbddtau=0$ (light gray curve) but
does not for $\alpha_0=\daddtau=0$ and $\beta_0=\dbddtau=1$ (dash-dot
curve).  In the former case of non-zero $\alpha$ with a double
peak, increasing $\beta_0$ enhances the stabilisation of the tensile
mechanism at $45^\circ$ and deepens the valley between low-angle and
high-angle peaks of $\dot{s}$ (Fig.~\ref{fig:beta}c--d). This occurs
because $\tau^{(1)}_{XX}$ is enhanced by the overall strengthening
effect of $\beta_0$ (Fig.~\ref{fig:beta}e). The very low band angles
that emerge in the numerical simulation with dynamic anisotropy and
$r_\beta=1$ (Fig.~\ref{fig:numerical}) are therefore a consequence of
both the dynamic effect of $\alpha$ and the enhancement by $\beta$.

\section{Summary and discussion}

We have developed and analysed a model of coupled magma/mantle
dynamics with anisotropic viscosity.  The anisotropy is controlled by
the orientation of principal stresses and the amount of deviatoric
stress. The model presented here introduces small modifications on
that of \cite{takei13}; in particular, the parameter $\beta$ models an
increase in contiguity of grains in the direction of maximum
compressive stress and the parameter $\tauoff$ allows for a finer
control on the magnitude of anisotropy and its sensitivity to stress
(for $r_{\beta}=\textrm{const.}$).  This description of viscous
anisotropy is physically consistent with experiments and relatively
simple, so its analysis should clarify the mechanics of rocks for
which the assumptions hold. Existing experimental data, however, are
not enough to quantitatively constrain all parameter values. The
parameter studies performed here aim to understand the underlying
physics.

It is known from previous theoretical work that anisotropic viscosity
lowers the angle of emergent, high-porosity bands.  Numerical
solutions (Fig.~\ref{fig:numerical}) compare uniform anisotropy
imposed \textit{a priori} with anisotropy that varies according to
local conditions of stress. They show that dynamic anisotropy leads to
lowering of band angle as compared with uniform anisotropy, where the
mean magnitude and angle from the dynamic case are used in the static
case. Moreover, dynamic anisotropy produces low-angle bands even when
its mean values wouldn't do so if applied uniformly and held constant.
The physical reasons for this have not previously been clear.  Indeed,
the question of why anisotropic viscosity lowers band angle at all has
not previously been addressed.

Static viscous anisotropy, in which viscous resistance to extension in
the most tensile direction is decreased, predicts low-angles of
high-porosity bands for two reasons: (\textit{a}) it suppresses the
mode of instability in which tension causes extension across
high-porosity bands; (\textit{b}) it creates a mode of instability in
which shear stress causes extension across high-porosity bands.  The
tensile instability has a peak perturbation growth rate in the maximum
tensile direction ($\theta=45^{\circ}$). When this instability is
suppressed by static anisotropy, the peak growth rate shifts to the
smaller angles that are favoured by the emergent shear instability.
And although the growth of the lowest angle bands are enhanced by the
shear instability, perturbations parallel to the shear plane
($\theta=0^{\circ}$) are stable because of the compressive stress
created by the base-state flow. Therefore, a low but finite angle of
high-porosity bands is predicted by this model.  Allowing for an
increase in contiguity and viscosity in the direction of maximum
compression has counter-balancing effects that leave predicted band
angles almost unchanged.

Dynamic viscous anisotropy, in which the anisotropy parameters are
allowed to vary with the local orientation and magnitude of deviatoric
stress, tends to further lower band angles.  It does so because it
suppresses the tensile instability around $\theta=45^{\circ}$ via the
following dynamic effect. Lower deviatoric stress in viscously weak
bands gives lower anisotropy there, which makes them less compliant to
tensile stress across them. Enhanced anisotropy in the interleaved,
lower-porosity regions makes those regions more compliant. This effect
over-compensates the compliance variations directly due to porosity
weakening; it favours melt segregation from the bands into the
inter-bands. Allowing for an increase in contiguity and viscosity in
the direction of maximum compressional stress increases the contrast
in band-parallel compressional stress (and deviatoric stress) between
bands and inter-bands.  This enhances the contrast in anisotropy and
further suppresses the tensile instability. Dynamic anisotropy makes
almost no modification to the shear instability.

The additional effects of dynamic anisotropy and the anisotropic
increase of contiguity are important because they make more robust the
prediction of low band angles.  Under static anisotropy, the mean
magnitude of anisotropy must be quite high to produce low-angle bands;
moderate levels are insufficient. In contrast, under dynamic
anisotropy with contiguity-increase in the direction of maximum
compression, moderate levels of mean anisotropy efficiently produce
low-angle bands.  This helps to support the hypothesis that low-angle
bands in experiments are due to anisotropic viscosity because it
expands the parameter space in which the theoretical predictions
should hold.

These conclusions were reached by use of stability analysis in a
coordinate system that is rotated with respect to the plane of simple
shear; in particular, the coordinate system is aligned with the
wavefronts of the harmonic perturbations.  This rotation leads to
simpler expressions for the growth rate of perturbations: the tensile
and shear modes appear as distinct terms that are amenable to physical
understanding. For this reason, our analysis represents a framework in
which to test and understand the family of rheologies that potentially
produce low-angle bands in shearing flows.  This includes variants of
isotropic and anisotropic viscosity, but also potentially of
dilatational granular rheology, damage, or composite rheologies
\citep[e.g.][]{rudge15}.

An application of the rotated coordinate system to the isotropic
power-law creep model with stress exponent $n$ \citep{katz06} is
presented in Appendix~\ref{sec:power-law}. As with all other models
considered here, porosity perturbations reduce viscosity in the bands,
resulting in the enhancement of the normal and shear strain rates,
$\strr^{(1)}_{YY}$ and $\strr^{(1)}_{XY}$. In this model, however, the
enhanced strain rates further reduce the viscosity, which feeds back
to further enhance the strain rates. The importance of this
non-Newtonian feedback relative to the porosity-weakening feedback is
roughly approximated by $n-1$. Both normal and shear strain rates
contribute to the non-Newtonian feedback; the relative importance of
the shear component increases with increasing $r_\xi$. Therefore, if
$n$ and $r_\xi$ are sufficiently large, shear strain rate is the key
weakening factor and the growth rate reaches a maximum at a
substantially lowered angle.  However, in contrast to anisotropic
viscosity, strain-rate weakening viscosity does not give rise to the
shear instability --- it merely lowers the most favorable angle for
the tensile instability (comparison in Tab.~\ref{table:summary}).
Although the details differ, both models predict an important role for
shear stress in the lowering mechanism; both predict a low but finite
angle with localised shear strain in the higher-porosity bands.

The model of viscous anisotropy used here seems promising as an
explanation for laboratory experiments on deformation of partially
molten rocks. Although its detailed form must be considered tentative,
we are not aware of another theory that reproduces the low-angle bands
found in experiments \citep{holtzman03a} while respecting the measured
stress-dependence of creep viscosity \citep{king10}. Furthermore,
radially inward migration of magma in experiments employing torsional
deformation \citep{king11a, qi15} may be direct evidence of base-state
segregation, a feature that arises naturally from viscous anisotropy
\citep{takei13} but may be impossible to reconcile with isotropic
viscosity. Although the present study focuses on the angle of bands,
the growth rate of bands is also affected by static and dynamic
anisotropy; the growth rate is lowered by static anisotropy and
further lowered by dynamic anisotropy (Figs.~\ref{fig:static} and
\ref{fig:dynamic}). This can be also discerned in the different total
strain and different ranges of porosity in Figs.~\ref{fig:numerical}a
and \ref{fig:numerical}b. Therefore, a quantitative comparison between
the measured and predicted growth rate becomes important for further
refining and testing the theory.

In the present theory, $\alpha$, $\beta$, and $\Theta$ are assumed to
depend on stress, based on the experimental results by \cite{daines97}
and \cite{takei10}. Although this assumption is considered to be valid
at small strain, possible evolution of these parameters with
increasing strain has to be investigated to model the system at large
strains. Indeed, for more than 200\% strain under simple shear,
\cite{zimmerman99} observed that the long axis of melt pockets is
predominantly oriented at an angle of $20^{\circ}$ from $\tau_1$; this
is difficult to explain by stress alone. It should be noted, however,
that microstructural analysis in laboratory studies has been performed
in terms of shape and orientation of melt pockets; an analysis in
terms of observed contiguity is more appropriate for comparison with
and incorporation into the model. Numerical simulation using dynamic
anisotropy and an empirically justified evolution equation for
contiguity will be important in future work.

\vspace{3mm}
\noindent\textbf{Acknowledgements}~~~The research leading to
these results has received funding from the European Research Council
(ERC) under the European Union’s Seventh Framework Programme
(FP7/2007–2013)/ERC grant agreement 279925. R.F.K.~visited the
Earthquake Research Institute of the University of Tokyo with support
from the International Research Promotion Office; he is grateful for
support by the Leverhulme Trust. Numerical simulations were performed
at the Advanced Research Computing facility of the University of
Oxford. The authors are grateful for stimulating discussions with
M.~Spiegelman, D.L.~Kohlstedt, and C.~Qi, and for helpful and
encouraging reviews by S.~Butler and two anonymous referees.

\appendix

\section{Power-law creep model by Katz \textit{et al.} (2006)}
\label{sec:power-law}

The model of band-formation under power-law viscosity by \cite{katz06}
is formulated by equations~\eqref{eq:governing} and the viscous
constitutive relations
\begin{equation}  
  C_{ijkl} = \eta(\phi, \strr_{II})　\times 
  \bordermatrix{ ij\!\downarrow &kl\!\rightarrow XX&YY&XY  \cr
    XX&r_\xi+\frac{4 }{3} & r_\xi-\frac{2}{3} &0 \cr
    YY& \cdot  & r_\xi+\frac{4 }{3} &0  \cr
    XY&\cdot &\cdot&1},
  \label{eq:CsystemK}
\end{equation}
where only 6 of the 16 components of the two-dimensional version are
shown due to the symmetry of $C_{ijkl}$. The normalised shear
viscosity $\eta(\phi, \strr_{II})$ depends on porosity and the second
invariant of the strain-rate tensor,
$\strr_{II} = \sqrt{\strr_{ij}\strr_{ij}/2}$, as
\begin{equation}
  \eta(\phi, \strr_{II}) = 
  \exp\left[-\lambda(\phi-\phi_0)/n\right] \strr_{II}^{\frac{1-n}{n}}
  \label{eq:eta_powlaw}
\end{equation}
\citep{katz06, takei09c}.  Equation \eqref{eq:eta_powlaw} represents a
power-law viscosity that, to represent deformation by dislocation
creep, has an exponent $n\approx3.5$ \citep[e.g.][]{karato93} ($n=1$
corresponds to Newtonian viscosity). Under dislocation creep, the
strain rate is highly sensitive to the stress because dislocation
velocity and density both increase with increasing stress. Hence the
model of \cite{katz06} incorporates strain-rate weakening in addition
to the porosity weakening.

\cite{katz06} demonstrated that strain-rate weakening viscosity works
to lower the band angle. Although the mechanism of this lowering is
briefly discussed in their paper, further analysis of their model
using the perturbation-oriented coordinate system is helpful to
understand their explanation and to compare it with the mechanism of
viscous anisotropy.  For consistency with the foregoing development,
$\eta_L=0$ is assumed here. We can expand~\eqref{eq:eta_powlaw} into
base-state and perturbation terms as
\begin{equation}
  \eta = \eta_0 
  \left\{1 - \verysmall
  \left[\frac{\lambda\phi_1}{n} +
    2\frac{n-1}{n}
    \left(2\strr^{(0)}_{XY}\strr^{(1)}_{XY}
      +\strr^{(0)}_{YY}\strr^{(1)}_{YY}\right)\right]\right\},
  \label{eq:eta1K}
\end{equation}
where
$\eta_0 = \left(\strr_{II}^{(0)}\right)^{\frac{1-n}{n}} =
2^{\frac{n-1}{n}}$
and we have used $\strr_{II}^{(0)} = 1/2$ and
$\strr^{(1)}_{XX}=0$. Combining \eqref{eq:CsystemK} and
\eqref{eq:eta1K} with stress balance \eqref{eq:eqmo3_1st2a} and
\eqref{eq:eqmo3_1st2b}, we obtain
\begin{subequations}
  \label{eq:static1K}
  \begin{align}
    \label{eq:band-normal}
    C^{(0)}_{YYYY}\strr^{(1)}_{YY}
    &= \left[\frac{\lambda}{n}\phi_1  + 2\frac{n-1}{n}\left(
      \strr^{(0)}_{YY}\strr^{(1)}_{YY} +
      2\strr^{(0)}_{XY}\strr^{(1)}_{XY}\right)\right]\tau^{(0)}_{YY}, \\
    \label{eq:band-shear}
    2C^{(0)}_{XYXY}\strr^{(1)}_{XY}
    &=\left[\frac{\lambda}{n}\phi_1 + 2\frac{n-1}{n} \left(
      \strr^{(0)}_{YY}\strr^{(1)}_{YY}
      + 2\strr^{(0)}_{XY}\strr^{(1)}_{XY}\right)\right]\tau^{(0)}_{XY}.
  \end{align}
\end{subequations}
This formulation is not the most amenable to inversion for the
strain-rate perturbations, but it allows for a clear comparison with
equations~\eqref{eq:strategy} and \eqref{eq:forcing}.  We have moved
terms to the right-hand side that can be considered to comprise the
forcing stresses $\tau^{(f)}_{YY}$ and $\tau^{(f)}_{XY}$. Two points
are evident: First, the forcing stresses retain the term representing
base-state stress operating on porosity perturbations. Second, there
are new terms that cross-couple the equations~\eqref{eq:static1K}.

The cross-coupling terms in \eqref{eq:static1K} arise because normal
$\strr_{YY}$ and shear $\strr_{XY}$ components both affect
$\strr_{II}$ and hence modify the viscosity (by way of an increase in
dislocation density). Two feedback mechanisms are thus at work,
causing growth of porosity perturbations. The first of these is a
direct effect: when $\lambda>0$, high-porosity bands are weaker by
virtue of their higher porosity. The second is indirect:
porosity-weakened bands have a larger strain-rate that, when $n>1$,
further weakens them through the non-linear viscosity. The relative
importance of the 2nd mechanism to the 1st one increases with
increasing $n-1$.

\begin{figure}
  \centering  \vspace{3mm}
  \includegraphics[width=8cm]{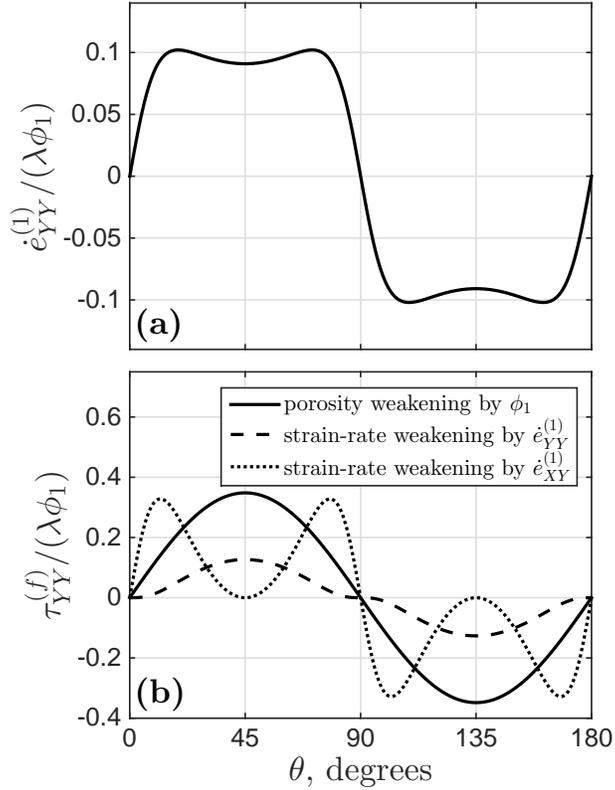}
  \caption{Characteristics of the power-law isotropic viscosity model
    as a function of the angle between porosity perturbations and the
    shear plane.  In both panels, $n=5$.  \textbf{(a)} Normal strain
    rate perturbation, which shows a double peak. \textbf{(b)} Forcing
    normal stress $\tau^{(f)}_{YY}$ due to porosity weakening (solid
    curve), strain-rate weakening associated with $\strr^{(1)}_{YY}$
    (dashed curve), and strain-rate weakening associated with
    $\strr^{(1)}_{XY}$ (dotted curve). The three curves in panel
    \textbf{(b)} sum to the curve in panel \textbf{(a)}.}
  \label{fig:powerlaw}
\end{figure}

Solving equations \eqref{eq:static1K} for $\strr^{(1)}_{YY}$ and
$\strr^{(1)}_{XY}$ and using equations~\eqref{eq:eqmo1_1st2} and
\eqref{eq:CsystemK}, the growth-rate is
\begin{equation}
  \dot{s}  = (1-\phi_0)\frac{\lambda}{n}\frac{\tau_{YY}^{(0)}}{C^{(0)}_{YYYY}}
  \left[1-4\frac{n-1}{n}\left(\strr^{(0)}_{XY}\right)^2
    -2\frac{n-1}{n}\left(1-\tfrac{r_\xi-2/3}{r_\xi+4/3}\right)
    \left(\strr^{(0)}_{YY}\right)^2\right]^{-1}.
  \label{eq:growthrate_katz}
\end{equation}
\cite{takei09c} obtained the identical result for $ \dot{s} $ and
showed that a single peak splits into two at large $n$. Here, to
understand the mechanism of the split, the forcing stress associated
with tension, $\tau^{(f)}_{YY}$, is plotted in
Figure~\ref{fig:powerlaw}b for each of the three terms in the RHS of
\eqref{eq:band-normal}.  Although the tensile forcing stress due to
the porosity and normal-strain rate perturbations are maximum at
$\theta=45^\circ$ (solid and dashed curves), that due to the
enhanced shear strain rate $\strr^{(1)}_{XY}$ has peaks at
$\theta\simeq 10^\circ$ and $80^\circ$ (dotted curve). The sum of
these three curves determines the profile of $\strr^{(1)}_{YY}$ in
panel~(a) and hence determines the growth-rate.  Panel~(b) confirms
that the weakening of viscosity by the enhanced shear-strain rate
$\strr^{(1)}_{XY}$ is the main cause of the peak split of the growth
rate.

\section{Calculation of the dynamic-anisotropy growth rate}
\label{sec:calculation-details}

To solve the first-order equations of force balance
\eqref{eq:first-order-simpler}, we need an expansion of the viscosity
tensor into its base-state and perturbation
components. Equation~\eqref{eq:Csystem} gives $C_{ijkl}$ as a function
of $\phi$, $\alpha$, $\beta$, and $\anisoangle$.  Under the dynamic
anisotropy model, it is necessary to account for non-zero
perturbations $\alpha_1$, $\beta_1$, and $\anisoangle_1$. In that
case, $C^{(0)}_{ijkl}$ and $C^{(1)}_{ijkl}$ are calculated as
\begin{subequations}
  \label{eq:visctensor_expansion}
  \begin{align}
    C^{(0)}_{ijkl}&=C_{ijkl}(\phi_0, \alpha_0, \beta_0, \anisoangle_0), \\
    C^{(1)}_{ijkl}&=\displaystyle
      \left.\pdiff{C_{ijkl}}{\phi}\right\vert_{0}\, \phi_1
       +   \left( \left.\pdiff{C_{ijkl}}{\alpha}\right\vert_{0}  
       +  r_\beta \left.\pdiff{C_{ijkl}}{\beta}\right\vert_{0}
                    \right)\alpha_1
       +   \left.\pdiff{C_{ijkl}}{\anisoangle}\right\vert_{0} \,\anisoangle_1. 
       \label{eq:C0C1}
  \end{align}
\end{subequations}
From the equation for anisotropy magnitude \eqref{eq:alpha_dynamic},
\begin{equation}
  \alpha_0 = 1 + \tanh\left(\frac{2\Delta \tau^{(0)} 
      -2\tauoff}{\tausat}\right),
  \label{eq:alpha0}
\end{equation}
where $\Delta \tau^{(0)} = 2(1 - (\alpha_0-\beta_0)/4)$.  Then from equation
\eqref{eq:C0C1}, the stress perturbation is written as
\begin{equation}
  \label{eq:stress_perturbation}
  \tau^{(1)}_{ij} = C^{(0)}_{ijkl} \strr^{(1)}_{kl} 
  -\lambda  \phi_1 \tau_{ij}^{(0)} 
  +\alpha_1\left( \left. \pdiff{C_{ijkl}}{\alpha}\right\vert_{0}+r_\beta
    \left.\pdiff{C_{ijkl}}{\beta}\right\vert_{0}\right)\strr^{(0)}_{kl}  
  + \anisoangle_1 \left.\pdiff{C_{ijkl}}
    {\anisoangle} \right\vert_{0}\strr^{(0)}_{kl}.
\end{equation}
Using \eqref{eq:stress_perturbation}, \eqref{eq:Csystem},
\eqref{eq:eij0}, and $\Theta_0 = \pi/4 + \theta$, the mechanical equilibrium
conditions $\tau^{(1)}_{XY}=\tau^{(1)}_{YY}=0$ from
equations~\eqref{eq:first-order-simpler} are written as
\begin{subequations}
  \label{eq:balance_dynamic}
  \begin{align}
    \label{eq:balance_dynamic_yyyy}
    C^{(0)}_{YYYY} \strr^{(1)}_{YY} + 2C^{(0)}_{YYXY} \strr^{(1)}_{XY}
    &= \lambda  \phi_1 \tau_{YY}^{(0)} + 
      \frac{1-r_\beta}{4}\left(\alpha_1\sin 2\theta
      +2\alpha_0\anisoangle_1 \cos 2\theta\right) 
      +\frac{1+r_\beta}{4}\alpha_1, \\[2mm] 
    C^{(0)}_{XYYY} \strr^{(1)}_{YY}  + 2C^{(0)}_{XYXY}\strr^{(1)}_{XY}
    &= \lambda  \phi_1\tau_{XY}^{(0)} + 
      \frac{1-r_\beta}{4}\left(\alpha_1\cos 2\theta
      - 2\alpha_0\anisoangle_1 \sin 2\theta\right),
  \end{align}
\end{subequations}
where the right-hand sides of these equations are the
dynamic-anisotropy version of the forcing stresses $\tau^{(f)}_{YY}$
and $\tau^{(f)}_{XY}$, respectively. The normal-stress perturbation in
the $X$-direction is
\begin{align}
  \tau^{(1)}_{XX} &= 
  C^{(0)}_{XXYY} \strr^{(1)}_{YY} + 2 C^{(0)}_{XXXY} \strr^{(1)}_{XY} 
  - \lambda  \phi_1 \tau_{XX}^{(0)} \nonumber\\
  &- \frac{1+r_\beta}{4}\alpha_1 
  + \frac{1-r_\beta}{4}\left(\alpha_1\sin 2\theta 
    + 2\alpha_0\anisoangle_1 \cos 2\theta\right).
  \label{eq:stressXX}
\end{align}

Microstructural anisotropy is determined by deviatoric stress.  From
the total differentials of equations~\eqref{eq:Theta} and
\eqref{eq:alpha_dynamic}, and from 
$\tau_{XY}^{(1)}=\tau_{YY}^{(1)}=0$, $\alpha_1$ and $\anisoangle_1$
are related to $\tau_{XX}^{(1)}$ as
\begin{subequations}
  \label{eq:alpha1_Theta1}
  \begin{align}
    \label{eq:alpha1}
    \alpha_1&=\daddtau \Delta \tau^{(1)}= \daddtau 
              \frac{\tau_{XX}^{(0)}-\tau_{YY}^{(0)}}{\Delta
              \tau^{(0)}}\tau_{XX}^{(1)},\\[2mm]
    \anisoangle_1&= - \frac{\sin 4\anisoangle_0}{4} 
                   \frac{\tau_{XX}^{(1)}}{\tau_{XX}^{(0)}-\tau_{YY}^{(0)}},
  \end{align}
\end{subequations}
with
\begin{equation}
\label{eq:alphacap}
  \daddtau = \left.\frac{\partial \alpha}
    {\partial \Delta \tau}\right|_{\Delta \tau=\Delta \tau^{(0)}}
  =\frac{2}{\tausat}\mbox{sech}^2
  \displaystyle\left(\frac{2(\Delta \tau^{(0)} - 
      \tauoff)}{\tausat}\right).
\end{equation}
Then we use the expression \eqref{eq:stressXX} for $\tau_{XX}^{(1)}$
and equations~\eqref{eq:alpha1_Theta1} to obtain
\begin{subequations}
  \label{eq:alpha1_Theta1_II}
  \begin{align}
    \label{eq:alpha1_II}
    \alpha_1&=2\zeta\left(
              C^{(0)}_{XXYY} \strr^{(1)}_{YY} + 2 C^{(0)}_{XXXY} \strr^{(1)}_{XY} 
              -\lambda  \phi_1 \tau_{XX}^{(0)}\right),\\
    \label{eq:Theta1_II}
    \anisoangle_1&=\frac{\zeta \Delta\tau^{(0)}\sin4\theta}
                   {2\daddtau  (\tau_{XX}^{(0)}-\tau_{YY}^{(0)} )^2}
                   \left(C^{(0)}_{XXYY} \strr^{(1)}_{YY} 
                   + 2 C^{(0)}_{XXXY} \strr^{(1)}_{XY} 
                   -\lambda  \phi_1 \tau_{XX}^{(0)}\right),
  \end{align}
\end{subequations}
where 
\begin{equation}
  \label{eq:zeta}
  \zeta^{-1}=\left(\frac{1+r_\beta}{2}-\frac{1-r_\beta}{2}\sin2\theta
  \right) + \frac{2\Delta\tau^{(0)}}{\daddtau
    (\tau_{XX}^{(0)}-\tau_{YY}^{(0)})} \left[ 1-
    \frac{(\alpha_0-\beta_0)}{8 (\tau_{XX}^{(0)}-\tau_{YY}^{(0)})}
    \cos 2\theta \sin 4\theta \right].
\end{equation}
Equations~\eqref{eq:alpha1_Theta1_II} can be substituted into the
stress-balance equations~\eqref{eq:balance_dynamic} giving a system in
which the only first-order quantities are $\phi_1$,
$\strr^{(1)}_{XY}$, and $\strr^{(1)}_{YY}$.

\bibliographystyle{abbrvnat}
\bibliography{manuscript}

\begin{thebibliography}{30}
\providecommand{\natexlab}[1]{#1}
\providecommand{\url}[1]{\texttt{#1}}
\expandafter\ifx\csname urlstyle\endcsname\relax
  \providecommand{\doi}[1]{doi: #1}\else
  \providecommand{\doi}{doi: \begingroup \urlstyle{rm}\Url}\fi

\bibitem[Allwright and Katz(2014)]{allwright14}
J.~Allwright and R.~Katz.
\newblock Pipe poiseuille flow of viscously anisotropic, partially molten rock.
\newblock \emph{Geophys.\ J.\ Int.}, 199\penalty0 (3):\penalty0 1608--1624,
  2014.
\newblock \doi{10.1093/gji/ggu345}.

\bibitem[Balay et~al.(2001)Balay, Buschelman, Gropp, Kaushik, Knepley, McInnes,
  Smith, and Zhang]{petsc-homepage}
S.~Balay, K.~Buschelman, W.~Gropp, D.~Kaushik, M.~Knepley, L.~McInnes,
  B.~Smith, and H.~Zhang.
\newblock http://www.mcs.anl.gov/petsc, 2001.

\bibitem[Balay et~al.(2004)Balay, Buschelman, Gropp, Kaushik, Knepley,
  {McInnes}, Smith, and Zhang]{petsc-manual}
S.~Balay, K.~Buschelman, W.~Gropp, D.~Kaushik, M.~Knepley, L.~{McInnes},
  B.~Smith, and H.~Zhang.
\newblock {PETSc} users manual.
\newblock Technical report, Argonne National Lab, 2004.

\bibitem[Butler(2012)]{butler12}
S.~Butler.
\newblock {Numerical Models of Shear-Induced Melt Band Formation with
  Anisotropic Matrix Viscosity}.
\newblock \emph{Phys.\ Earth Planet.\ In.}, 200-201:\penalty0 28--36, 2012.
\newblock \doi{10.1016/j.pepi.2012.03.011}.

\bibitem[Cooper et~al.(1989)Cooper, Kohlstedt, and Chyung]{cooper89}
R.~Cooper, D.~Kohlstedt, and K.~Chyung.
\newblock Solution-precipitation enhanced creep in solid–liquid aggregates
  which display a non-zero dihedral angle.
\newblock \emph{Acta Metall}, 37:\penalty0 1759--1771, 1989.

\bibitem[Daines and Kohlstedt(1997)]{daines97}
M.~Daines and D.~Kohlstedt.
\newblock Influence of deformation on melt topology in peridotites.
\newblock \emph{J.\ Geophys.\ Res.}, 102:\penalty0 10257--10271, 1997.

\bibitem[Drew(1983)]{drew83}
D.~Drew.
\newblock {Mathematical modeling of two-phase flow}.
\newblock \emph{Annual Review Of Fluid Mechanics}, 15:\penalty0 261--291, 1983.
\newblock \doi{10.1146/annurev.fl.15.010183.001401}.

\bibitem[Holtzman and Kohlstedt(2007)]{holtzman07}
B.~Holtzman and D.~Kohlstedt.
\newblock Stress-driven melt segregation and strain partitioning in partially
  molten rocks: Effects of stress and strain.
\newblock \emph{J.\ Petrol.}, 48:\penalty0 2379--2406, 2007.
\newblock \doi{10.1093/petrology/egm065}.

\bibitem[Holtzman et~al.(2003)Holtzman, Groebner, Zimmerman, Ginsberg, and
  Kohlstedt]{holtzman03a}
B.~Holtzman, N.~Groebner, M.~Zimmerman, S.~Ginsberg, and D.~Kohlstedt.
\newblock Stress-driven melt segregation in partially molten rocks.
\newblock \emph{Geochem.\ Geophys.\ Geosys.}, 4, 2003.
\newblock \doi{10.1029/2001GC000258}.

\bibitem[Karato and Wu(1993)]{karato93}
S.~Karato and P.~Wu.
\newblock Rheology of the upper mantle - a synthesis.
\newblock \emph{Science}, 260, 1993.

\bibitem[Katz and Takei(2013)]{katz13}
R.~Katz and Y.~Takei.
\newblock {Consequences of viscous anisotropy in a deforming, two-phase
  aggregate: 2. Numerical solutions of the full equations}.
\newblock \emph{J.\ Fluid\ Mech.}, 734:\penalty0 456--485, 2013.
\newblock \doi{10.1017/jfm.2013.483}.

\bibitem[Katz et~al.(2006)Katz, Spiegelman, and Holtzman]{katz06}
R.~Katz, M.~Spiegelman, and B.~Holtzman.
\newblock The dynamics of melt and shear localization in partially molten
  aggregates.
\newblock \emph{Nature}, 442, 2006.
\newblock \doi{10.1038/nature05039}.

\bibitem[Katz et~al.(2007)Katz, Knepley, Smith, Spiegelman, and Coon]{katz07}
R.~Katz, M.~Knepley, B.~Smith, M.~Spiegelman, and E.~Coon.
\newblock Numerical simulation of geodynamic processes with the {P}ortable
  {E}xtensible {T}oolkit for {S}cientific {C}omputation.
\newblock \emph{Phys.\ Earth Planet.\ In.}, 163:\penalty0 52--68, 2007.
\newblock \doi{10.1016/j.pepi.2007.04.016}.

\bibitem[King et~al.(2010)King, Zimmerman, and Kohlstedt]{king10}
D.~King, M.~Zimmerman, and D.~Kohlstedt.
\newblock Stress-driven melt segregation in partially molten olivine-rich rocks
  deformed in torsion.
\newblock \emph{J.\ Petrol.}, 51:\penalty0 21--42, 2010.
\newblock \doi{10.1093/petrology/egp062}.

\bibitem[King et~al.(2011)King, Holtzman, and Kohlstedt]{king11a}
D.~S.~H. King, B.~K. Holtzman, and D.~L. Kohlstedt.
\newblock {An experimental investigation of the interactions between
  reaction-driven and stress-driven melt segregation: 1. Application to mantle
  melt extraction}.
\newblock \emph{Geochemistry Geophysics Geosystems}, 12, 2011.
\newblock \doi{10.1029/2011GC003684}.

\bibitem[Mc{K}enzie(1984)]{mckenzie84}
D.~Mc{K}enzie.
\newblock The generation and compaction of partially molten rock.
\newblock \emph{J.\ Petrol.}, 25, 1984.

\bibitem[Mei et~al.(2002)Mei, Bai, Hiraga, and Kohlstedt]{mei02}
S.~Mei, W.~Bai, T.~Hiraga, and D.~Kohlstedt.
\newblock Influence of melt on the creep behavior of olivine-basalt aggregates
  under hydrous conditions.
\newblock \emph{Earth Plan.\ Sci.\ Lett.}, 201:\penalty0 491--507, 2002.

\bibitem[Qi et~al.(2015)Qi, Kohlstedt, Katz, and Takei]{qi15}
C.~Qi, D.~Kohlstedt, R.~Katz, and Y.~Takei.
\newblock An experimental test of the viscous anisotropy hypothesis for
  partially molten rocks.
\newblock \emph{Proc.\ Nat.\ Acad.\ Sci.}, 2015.
\newblock \doi{10.1073/pnas.1513790112}.

\bibitem[Rudge and Bercovici(2015)]{rudge15}
J.~Rudge and D.~Bercovici.
\newblock {Melt-band instabilities with two-phase damage}.
\newblock \emph{Geophys.\ J.\ Int.}, 201\penalty0 (2):\penalty0 640--651, 2015.
\newblock \doi{10.1093/gji/ggv040}.

\bibitem[Rudge et~al.(2011)Rudge, Bercovici, and Spiegelman]{rudge11}
J.~F. Rudge, D.~Bercovici, and M.~Spiegelman.
\newblock {Disequilibrium melting of a two phase multicomponent mantle}.
\newblock \emph{Geophys.\ J.\ Int.}, 184\penalty0 (2):\penalty0 699--718, 2011.

\bibitem[Simpson et~al.(2010{\natexlab{a}})Simpson, Spiegelman, and
  Weinstein]{simpson10a}
G.~Simpson, M.~Spiegelman, and M.~Weinstein.
\newblock {A multiscale model of partial melts: 1. Effective equations}.
\newblock \emph{Journal Of Geophysical Research}, 115, 2010{\natexlab{a}}.
\newblock \doi{10.1029/2009JB006375}.

\bibitem[Simpson et~al.(2010{\natexlab{b}})Simpson, Spiegelman, and
  Weinstein]{simpson10b}
G.~Simpson, M.~Spiegelman, and M.~Weinstein.
\newblock {A multiscale model of partial melts: 2. Numerical results}.
\newblock \emph{Journal Of Geophysical Research}, 115, 2010{\natexlab{b}}.
\newblock \doi{10.1029/2009JB006376}.

\bibitem[Spiegelman(2003)]{spiegelman03b}
M.~Spiegelman.
\newblock Linear analysis of melt band formation by simple shear.
\newblock \emph{Geochem.\ Geophys.\ Geosys.}, 2003.
\newblock \doi{10.1029/2002GC000499}.

\bibitem[Stevenson(1989)]{stevenson89}
D.~Stevenson.
\newblock Spontaneous small-scale melt segregation in partial melts undergoing
  deformation.
\newblock \emph{Geophys.\ Res.\ Letts.}, 16, 1989.

\bibitem[Takei(1998)]{takei98}
Y.~Takei.
\newblock {Constitutive mechanical relations of solid-liquid composites in
  terms of grain-boundary contiguity}.
\newblock \emph{Journal Of Geophysical Research}, 103:\penalty0 18183--18203,
  1998.

\bibitem[Takei(2010)]{takei10}
Y.~Takei.
\newblock {Stress-induced anisotropy of partially molten rock analogue deformed
  under quasi-static loading test}.
\newblock \emph{Journal Of Geophysical Research}, 115:\penalty0 B03204, 2010.
\newblock \doi{10.1029/2009JB006568}.

\bibitem[Takei and Holtzman(2009{\natexlab{a}})]{takei09a}
Y.~Takei and B.~Holtzman.
\newblock Viscous constitutive relations of solid-liquid composites in terms of
  grain boundary contiguity: 1. {G}rain boundary diffusion control model.
\newblock \emph{J.\ Geophys.\ Res.}, 2009{\natexlab{a}}.
\newblock \doi{10.1029/2008JB005850}.

\bibitem[Takei and Holtzman(2009{\natexlab{b}})]{takei09c}
Y.~Takei and B.~Holtzman.
\newblock Viscous constitutive relations of solid-liquid composites in terms of
  grain boundary contiguity: 3. causes and consequences of viscous anisotropy.
\newblock \emph{J.\ Geophys.\ Res.}, 2009{\natexlab{b}}.
\newblock \doi{10.1029/2008JB005852}.

\bibitem[Takei and Katz(2013)]{takei13}
Y.~Takei and R.~Katz.
\newblock {Consequences of viscous anisotropy in a deforming, two-phase
  aggregate: 1. Governing equations and linearised analysis}.
\newblock \emph{J.\ Fluid\ Mech.}, 734:\penalty0 424--455, 2013.
\newblock \doi{10.1017/jfm.2013.482}.

\bibitem[Zimmerman et~al.(1999)Zimmerman, Zhang, Kohlstedt, and
  Karato]{zimmerman99}
M.~Zimmerman, S.~Zhang, D.~Kohlstedt, and S.~Karato.
\newblock {Melt distribution in mantle rocks deformed in shear}.
\newblock \emph{Geophys.\ Res.\ Letts.}, 26\penalty0 (10):\penalty0 1505--1508,
  1999.

\end{thebibliography}

\end{document}